\documentclass[12pt,draftclsnofoot,onecolumn]{IEEEtran}
\usepackage{amsmath}
\usepackage{amssymb}
\usepackage{amsfonts}
\usepackage{graphicx}
\usepackage{enumerate}
\usepackage{bbm}
\usepackage{subfig}
\usepackage{mathtools} 
\usepackage{cite}
\usepackage{url}
\usepackage{array}
\usepackage{verbatim}
\usepackage{comment}
\usepackage{stfloats}
\usepackage[pdfencoding=auto]{hyperref}

\usepackage{hhline}
\newcommand*\dd{\mathop{}\!\mathrm{d}}

\newtheorem{theorem}{Theorem}

\DeclareMathOperator*{\argmax}{argmax}

\usepackage{color}
\definecolor{red}{rgb}{1,0,0}
\definecolor{blue}{rgb}{0,0,1}

\usepackage[font=small,skip=-1pt]{caption}
\IEEEoverridecommandlockouts
\begin{document}
\title{Non-Orthogonal Multiple Access for mmWave Drone Networks with Limited Feedback}
\author{Nadisanka~Rupasinghe,~Yavuz~Yap{\i}c{\i},~\.{I}smail~G\"{u}ven\c{c}~and~Yuichi~Kakishima
\thanks{N.~Rupasinghe,~Y.~Yap{\i}c{\i},~and~\.{I}.~G\"{u}ven\c{c} are with the Department of Electrical and Computer Engineering, North Carolina State University, Raleigh, NC, 27606 (e-mail:~\{rprupasi,yyapici,iguvenc\}@ncsu.edu).

Yuichi Kakishima is with DOCOMO Innovations, Inc., Palo Alto, CA, 94304  (e-mail: kakishima@docomoinnovations.com).

This research was supported in part by U.S. National Science Foundation under the grant CNS-1618692.
}}%
\maketitle

\vspace{-4mm}

\begin{abstract}
Unmanned aerial vehicle (UAV) base stations (BSs) can be a promising solution to provide connectivity and quality of service (QoS) guarantees during temporary events and after disasters. In this paper, we consider a scenario where UAV-BSs are serving large number of mobile users in a hot spot area (e.g., in a stadium). We introduce non-orthogonal multiple access (NOMA) transmission at UAV-BSs to serve more users simultaneously considering user distances as the available feedback for user ordering during NOMA formulation. With millimeter-wave (mmWave) transmission and multi-antenna techniques, we assume UAV-BS generates directional beams and multiple users are served simultaneously within the same beam. However, due to the limitations of physical vertical beamwidth of the UAV-BS beam, it may not be possible to cover the entire user region at UAV altitudes of practical relevance. During such situations, a beam scanning approach is proposed to maximize the achievable sum rates. We develop a comprehensive framework over which outage probabilities and respective sum rates are derived rigorously and we investigate the optimal operational altitude of UAV-BS to maximize the sum rates using our analytical framework. Our analysis shows that NOMA with distance feedback can provide better outage sum rates compared to orthogonal multiple access.

\end{abstract}

\begin{IEEEkeywords}
$5$G, drone, HPPP, mmWave, non-orthogonal multiple access (NOMA), stadium, UAV.
\end{IEEEkeywords}

\section{Introduction}

Use of unmanned aerial vehicles (UAVs) as mobile base stations (BSs) can facilitate rapid deployment of a wireless network infrastructure during temporary events \cite{Att2, Times, BBC, Zeng2016UAVOpportunities,merwaday2016improved,
UAV_NOMA_Asilomar}. Such temporary events include natural disasters where the existing network infrastructure is destroyed \cite{Att2, merwaday2016improved}, or sports events in stadiums where there may be thousands of mobile users straining the available communication resources \cite{BBC, UAV_NOMA_Asilomar}. In such scenarios, achieving high spectral efficiency (SE) becomes critical to maintain the quality-of-service (QoS) requirements of all mobile users. To improve the SE, non-orthogonal multiple access (NOMA) is identified as a promising technology for next generation wireless communication systems~\cite{Saito13VTC,3GPP16NOMA_LTE}.
In contrast to the conventional orthogonal multiple access (OMA) schemes (e.g., time-division multiple access (TDMA)), NOMA simultaneously serves multiple users in non-orthogonal resources (in time, frequency, code and space domains)
by separating the users in the power domain. Therefore, it is a  suitable technology for effectively serving large number of wireless users, which is the case in large stadiums~\cite{Ding17PoorRandBeamforming,UAV_NOMA_Asilomar}  while enhancing SE. 

There are several recent use cases and many research efforts in academia which consider UAVs as aerial BSs. For example, AT\&T had recently deployed their flying COW (Cell on Wings) to provide data, voice, and text services to users in Puerto Rico in the aftermath of hurricane Maria~\cite{Att2, Times}. Nokia and British mobile operator EE are also considering drone BSs to provide coverage during emergency situations \cite{BBC}. AT\&T has been recently exploring the possibility of deploying UAV-BSs for augmenting their network capacity especially in hot spot scenarios~\cite{BBC}.

Further, there are several research studies in the academic literature considering UAVs as aerial BSs~\cite{merwaday2016improved, Abhay2017HetNetUAV, Nadisanka16GC,  Mozaffari2016UAVsOptimalCoverage, Mozaffari2015GC, Mozaffari2016UAVsPowerEffecient,  Bor2016UAVsEfficient3DPlacement, UAV_Placement_3, Flying_Drones}. When UAVs serving as aerial BSs in a communication network they can be deployed to achieve different requirements. By optimizing UAV locations through brute force searching techniques, \cite{merwaday2016improved} discusses achievable throughput coverage and $5$th percentile SE gains from the deployment of UAV-BSs during disaster situations where existing fixed communication infrastructure is damaged. In a follow up work \cite{Abhay2017HetNetUAV}, further enhanced inter-cell interference coordination (FeICIC) from 3GPP Release~$11$ is introduced along with genetic algorithm based UAV hovering optimization technique to improve achievable $5$th percentile SE further. In \cite{Nadisanka16GC}, multi-antenna techniques are exploited to optimize hovering locations of the UAV-BSs focusing on minimizing interference leakage and maximizing desired user signal-to-noise ratio (SNR). In \cite{Mozaffari2016UAVsOptimalCoverage}, by considering circle packing theory concepts, 3-Dimensional (3D) locations for deploying UAV-BSs with the objective of maximizing total coverage area is discussed. Impact of UAVs' altitude on the minimum required transmit power to achieve maximum ground coverage is analyzed in \cite{Mozaffari2015GC} for the case of two UAVs. An approach to identify hovering locations for UAVs in order to achieve power efficient DL transmission while satisfying the users' rate requirements is discussed in \cite{Mozaffari2016UAVsPowerEffecient}. In \cite{Bor2016UAVsEfficient3DPlacement}, a method to identify 3D hovering locations for UAV-BSs is proposed with the objective of maximizing the revenue of the network which is measured considering the number of users covered by the UAV-BS.

Use of NOMA techniques to improve SE have also been studied extensively in the literature in a broader context. In \cite{Saito13VTC}, a system-level performance evaluation based on 3GPP settings is carried out to identify potential performance gains with NOMA over orthogonal frequency division multiple access (OFDMA). Achievable performance with NOMA when users are randomly deployed is investigated in \cite{Ding14NOMAfor5G_RandUsers} considering two different criteria: 1) when each user has a targeted rate based on their QoS requirements; and 2) opportunistic user rates based on their channel conditions. Provided that the system parameters are appropriately chosen, better rate performance can be observed with NOMA compared to its orthogonal counterpart under both criteria. In \cite{Timotheou15PAforFairness}, a power allocation strategy for NOMA transmission is discussed considering user fairness in the downlink (DL) data transmission. In \cite{Ding16MIMO_NOMA}, multiple-input-multiple-output (MIMO) techniques are introduced to NOMA transmission along with user pairing and power allocation strategies to enhance MIMO-NOMA performance over MIMO-OMA. A general MIMO-NOMA framework applicable to both DL and uplink (UL) transmission is discussed in \cite{Ding16Schober_MIMO_NOMA} by considering signal alignment concepts. A cooperative NOMA transmission strategy for MIMO systems is proposed in \cite{Coop_NOMA_MIMO}. In that, a central user with strong channel condition acts as a relay for cell-edge user with poor channel condition to enhance its reception reliability. To protect from possible eavesdropper attacks, a secure beamforming strategy for NOMA transmission in the DL of multiple-input-single-output (MISO) systems is proposed in \cite{Secure_BF_NOMA}. In \cite{Ding17PoorRandBeamforming}, a random beamforming approach for millimeter (mmWave) NOMA networks is presented. To achieve power domain user separation, effective channel gains of users which depend on the angle offset between the randomly generated BS beam and user locations are considered. Two users are then served simultaneously within a single BS beam by employing NOMA techniques. A UAV based mobile cloud computing system is proposed in \cite{7932157} where UAVs offer computation offloading opportunities to mobile stations (MS) with limited local processing capabilities. In that, just for offloading purposes between UAV and MSs, NOMA is proposed as one viable solution. In our earlier work \cite{UAV_NOMA_Asilomar}, NOMA transmission is introduced to UAVs acting as aerial BSs to provide coverage over a stadium. Assuming the availability of full channel state information (CSI) for NOMA user ordering, some intuitive results based on computer simulations are provided for achievable rates without providing rigorous analytical evaluations.

In this paper, we consider a dense multi-user scenario with large number of mobile users (such as a stadium or a concert area), and a UAV-BS is deployed to provide coverage to those users. In order to improve spectral efficiency, we consider NOMA transmission specifically considering user distances as the available feedback and hence distance based user ordering. Our specific contributions can be summarized as follows.
\begin{itemize}
\item[i.] We first study the problem of UAV-BS coverage at a given sector of the stadium. To quantify the portion of the stadium sector that can be covered by a UAV-BS beam, we utilize the dimensions of the region where users are distributed 
(i.e., the \emph{user region}), altitude of the UAV-BS,  and the vertical beamwidth of the antenna propagation pattern. We show that the required vertical beamwidth to cover the entire user region is a concave function of the altitude, and, hence, the \emph{radiated region} by a UAV-BS beam only partially covers user region for a range of altitudes of practical relevance for UAV operations.
\item[ii.] Subsequently, a beam scanning approach is proposed to identify the best radiated region when the user region is covered only partially by a UAV-BS beam. We identify all unique events that correspond to presence of desired users for NOMA transmission within the physically radiated region, and formulate a hybrid transmission strategy serving all or part of those desired users simultaneously based on their existence within the radiated region.
\item[iii.] We consider NOMA transmission at the UAV-BS to increase spectral efficiency, and propose a practical feedback mechanism leveraging the user distance information (as partial CSI), which has a potential to reduce overall complexity with respect to the full CSI feedback approach as in \cite{UAV_NOMA_Asilomar} especially for rapidly fluctuating time-varying channels. For each user participating in NOMA, a target rate based on their QoS requirements is defined. We develop a comprehensive analytical framework over which outage probabilities and respective sum rates are derived rigorously for this distance feedback mechanism, which are verified through extensive simulations. We further provide a rigorous asymptotic analysis of the outage probabilities to obtain further insight. We investigate the optimal operational altitude of UAV-BS to maximize the sum rates using the developed analytical framework.
\end{itemize}

The rest of the paper is organized as follows. Section~\ref{sec:Sys_Model} captures the system model involving the NOMA transmission within a single UAV-BS DL beam. Formulation of NOMA transmission strategy is discussed in Section~\ref{Sec:NOMA_Transmission}, while outage probabilities and sum rates of NOMA with distance feedback are analytically investigated in Section~\ref{Sec:NOMARates_Distance_FB}. The respective numerical results are presented in Section~\ref{sec:Numerical_results}, and the paper ends with some concluding remarks in Section~\ref{sec:conclusion}.


\section{System Model} \label{sec:Sys_Model}

\begin{figure}[!t]
\begin{center}
\includegraphics[width=0.75\textwidth]{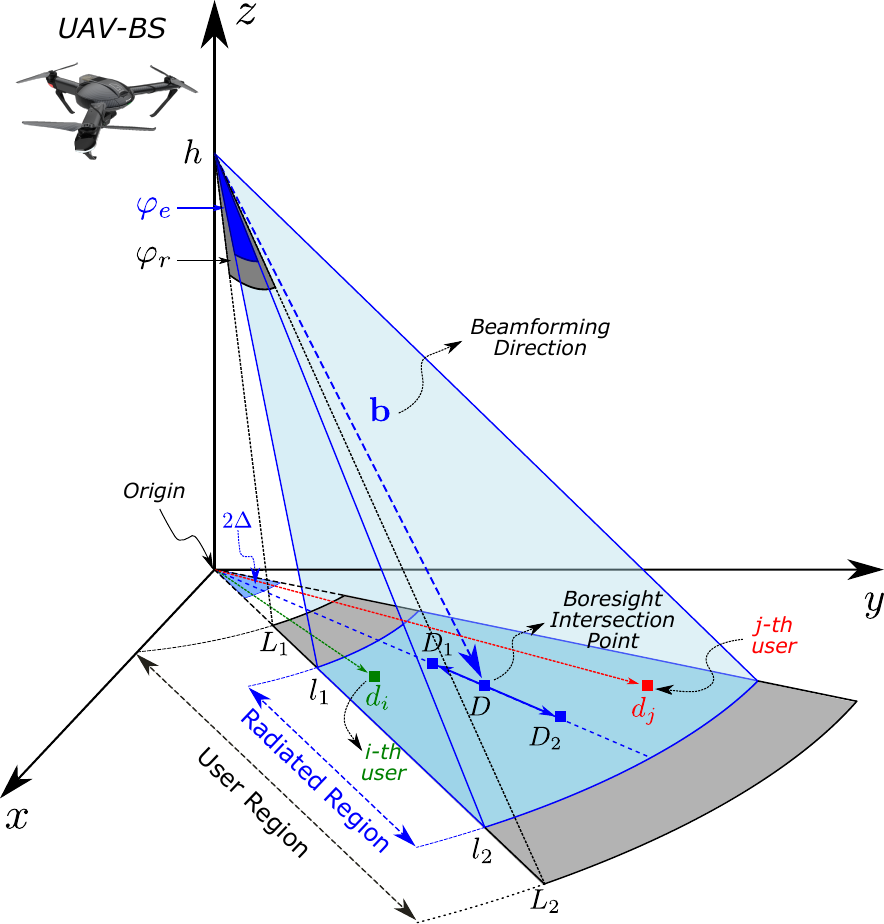}
\end{center}
\caption{A multiuser DL scenario, where the \emph{user region} is partially covered by the \textit{radiated region}. NOMA transmission serves $i$-th and $j$-th users simultaneously within a single DL beam.}
\label{fig:footprint}
\vspace{0.25 em}
\end{figure}

We consider a mmWave-NOMA transmission scenario where a single UAV-BS, which is equipped with an $M$~element antenna array, is serving mobile users in the DL. We assume that all these single-antenna users lie inside a specific \emph{user region} as shown in Fig.~\ref{fig:footprint}, and are represented by the index set $\mathcal{N}_{\rm U} = \{1,2,\ldots K\}$. We also assume that the user region of interest may or may not be fully covered by a 3-dimensional (3D) beam generated by the UAV-BS depending on the specific geometry of the environment, hovering altitude of the UAV-BS, and 3D antenna radiation pattern. In the specific scenario sketched in Fig.~\ref{fig:footprint}, only a smaller portion of the user region is being covered by the UAV-BS beam, and, hence, is labeled as \textit{radiated region}. The user region is identified by the inner-radius $L_1$, the outer-radius $L_2$, and $2\Delta$, which is the fixed angle within the projection of horizontal beamwidth of the UAV-BS antenna pattern on the $xy$-plane, as shown in Fig.~\ref{fig:footprint}. Similarly, the radiated region is described by the inner-radius $l_1$, outer-radius $l_2$, and angle $2\Delta$. Note that it is possible to reasonably model various different hot spot scenarios such as stadium, concert hall, traffic jam, and urban canyon by modifying these control parameters. For example, larger $L_1$ may correspond to a sports event where users are only allowed to use the available seats in the tribunes while smaller $L_1$ may represent a music concert event where users can also be present on the ground as well as the tribunes.

\subsection{User Distribution and mmWave Channel Model}
We assume that mobile users are randomly distributed following a homogeneous Poisson point process (HPPP) with density $\lambda$ \cite{Haenggi05Stochastic_Geo,merwaday2014capacity}. Hence, the number of users in the specified user region is Poisson distributed such that $\textrm{P}(K \textrm{ users in the user region})\,{=}\, \frac{\mu^K e^{{-}\mu}}{K!} $ with $\mu\,{=}\,(L_2^2\,{-}\,L_1^2)\Delta \lambda$. The channel $\textbf{h}_k$ between the $k$-th user in the user region and the UAV-BS is given as
\begin{align} \label{k_UE_original_channel}
\textbf{h}_k = \sqrt{M} \sum \limits _{p=1}^{N_{\rm P}} \frac{\alpha_{k,p} \textbf{a}(\theta_{k,p})}{\sqrt{\textrm{PL}\left(\sqrt{d_k^2 + h^2}\right)}},
\end{align} where $N_{\rm P}$, $h$, $d_k$, $\alpha_{k,p}$ and $\theta_{k,p}$ represent the number of multi-paths, UAV-BS hovering altitude, horizontal distance between $k$-th user and UAV-BS, gain of the $p$-th path which is complex Gaussian distributed with $\mathcal{CN}(0,1)$, and angle-of-departure (AoD) of the $p$-th path, respectively. $\textbf{a}(\theta_{k,p})$ is the steering vector with AoD $\theta_{k,p}$ given for uniform linear array (ULA) as \begin{align}\label{eqn:steering_vector}{
\textbf{a}\left( \theta_{k,p} \right) = \frac{1}{\sqrt{M}} \left[ 1 \;\; e^{-j2\pi \frac{D}{\zeta}\sin\left( \theta_{k,p}\right) } \; \dots \; e^{-j2\pi \frac{D}{\zeta}\left( M-1\right)\sin\left( \theta_{k,p} \right) } \right]^{\rm T} ,}
\end{align}
where $D$ is the antenna spacing in the ULA, and $\zeta$ is the wavelength. The path loss (PL) between $k$-th mobile user and the UAV-BS is captured by $\textrm{PL}\left(\sqrt{d_k^2 + h^2}\right)$. We can safely assume that all the users have line-of-sight (LoS) paths since UAV-BS is hovering at relatively high altitudes and the probability of having scatterers around UAV-BS is very small. In particular, as presented in \cite{ mmWave_Channel_Ch_Modeling_2, mmWave_Channel_Ch_Modeling_1}, the gains of NLoS paths are typically $20$~dB weaker than that of LoS path in mmWave channels. Hence, as considered in \cite{Lee16mmWaveChannel, mmWave_Channel_Cap_Analysis, mmWave_Channel_Cov_Analysis}, it is reasonable to assume a single LoS path for the mmWave channel under consideration, and \eqref{k_UE_original_channel} accordingly becomes
\begin{align} \label{k_UE_modified_LoS_channel}
\textbf{h}_k = \sqrt{M} \frac{\alpha_k \textbf{a}(\theta_k)}{\sqrt{\textrm{PL}\left(\sqrt{d_k^2 + h^2}\right)}},
\end{align} where $\theta_{k}$ is AoD of the LoS path.

\subsection{Coverage of User Region} \label{sec:user_region_coverage}
\begin{figure}[!t]
\begin{center}
\includegraphics[width=0.7\textwidth]{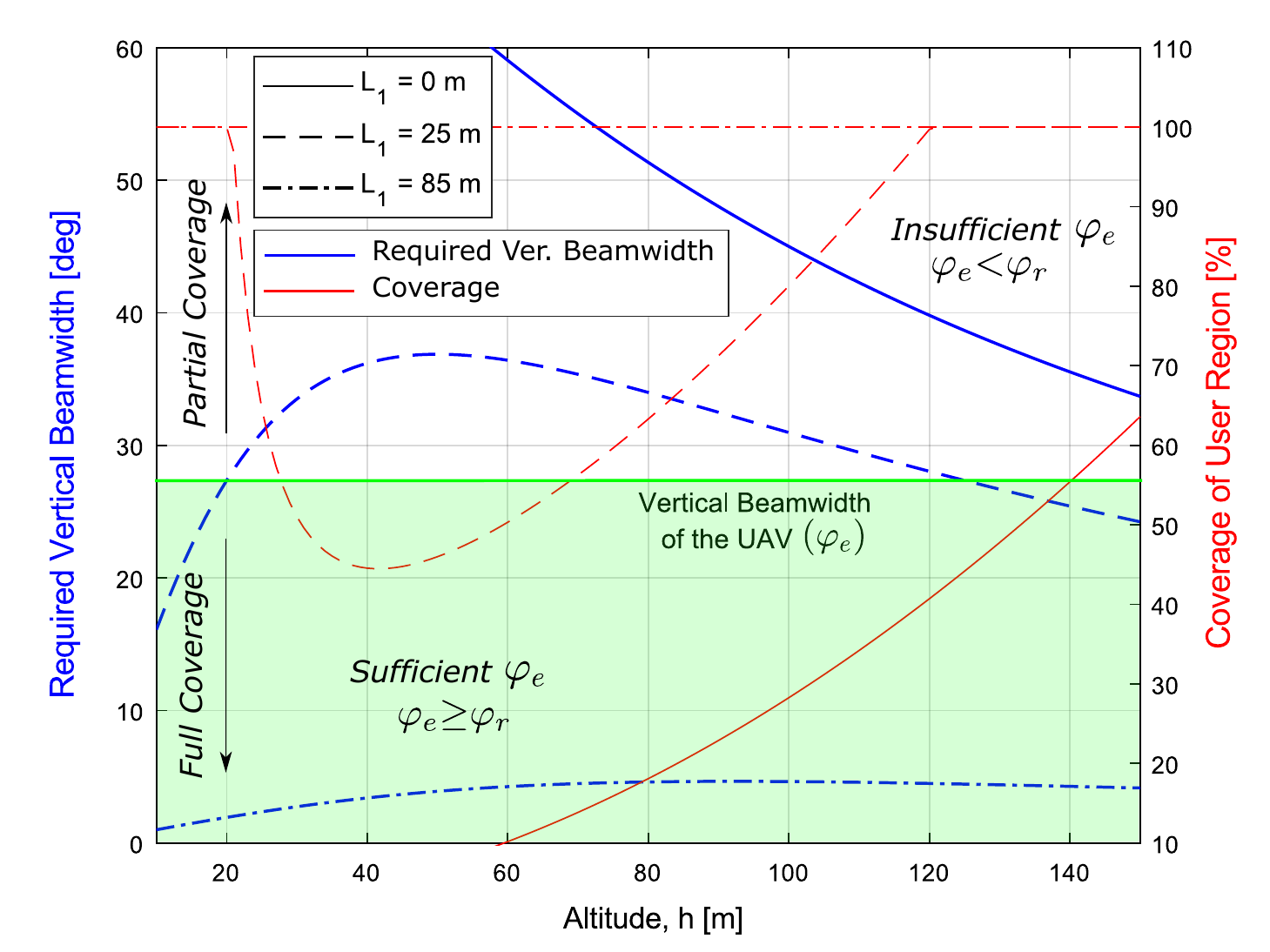}
\end{center}
\caption{Required vertical beamwidth $\varphi_r$ to cover the entire user region, and percentage of the user region covered by the UAV with vertical beamwidth of $\varphi_e\,{=}\, 28^{\circ}$ are depicted along with varying UAV altitude and user deployment choices ($L_1\,{=}\,\{0,25,85\}$~m, $L_2\,{=}\,100$~m). The upper and lower parts denote \textit{insufficient $\varphi_e$} ($\varphi_e\,{<}\,\varphi_r$) and \textit{sufficient $\varphi_e$} ($\varphi_e\,{\geq}\,\varphi_r$) regions, respectively, with respect to $\varphi_r$, representing partially/fully covered user region.
}
\label{fig:VertAngle}
\end{figure}

Due to the limited vertical beamwidth $\varphi_e$ of the UAV-BS antenna radiation pattern, it may not be possible to cover the entire user region all the time, as demonstrated by Fig.~\ref{fig:footprint}. To gain more insight into this \emph{partial coverage} problem, we plot the required vertical beamwidth $\varphi_r$ in Fig.~\ref{fig:VertAngle}, which shows the vertical beamwidth to cover the user region \textit{entirely} as a function of the UAV-BS altitude. Considering three different inner radius $L_1$ values while keeping the outer radius $L_2$ the same (see Fig.~\ref{fig:footprint} for relative geometry), we demonstrate the variation of $\varphi_r$ along with the UAV-BS altitude. In the right vertical axis of Fig.~\ref{fig:VertAngle}, we also show the percentage of the area covered within the user region by the UAV-BS beam, with a vertical beamwidth of $\varphi_e\,{=}\, 28^{\circ}$. We observe from Fig.~\ref{fig:VertAngle} that this vertical beamwidth $\varphi_e\,{=}\, 28^{\circ}$ (shown as a horizontal line) divides the required vertical beamwidth $\varphi_r$ space into two regions. In particular, the region denoted as \textit{insufficient~$\varphi_e$} in the upper part of Fig.~\ref{fig:VertAngle} corresponds to the case where $\varphi_e\,{<}\,\varphi_r$ and hence user region is covered only partially. On the other hand, the region denoted as \textit{sufficient $\varphi_e$} in the lower part of Fig.~\ref{fig:VertAngle} captures the case where user region is fully covered with $\varphi_e\,{\geq}\,\varphi_r$.

When the users are located everywhere in the stadium corresponding to the setting $L_1\,{=}\,0$~m, the required vertical angle $\varphi_r$ is very large and decreases monotonically with increasing UAV altitude, as depicted in Fig.~\ref{fig:VertAngle}. This intuitive result shows that the practical values of vertical beamwidth $\varphi_e$ can not cover the entire user region within the considered UAV altitude range. By setting the inner-radius $L_1\,{=}\,25$~m, it is possible to represent a scenario involving a big stage in the middle of the stadium during a music concert. In this case, the required vertical angle $\varphi_r$ to cover the entire user region first increases with UAV altitude, and then starts to decrease, as can be observed from Fig.~\ref{fig:VertAngle}. Because of this non-monotonic behavior of $\varphi_r$, practical values of $\varphi_e$ may not be sufficiently large to cover the entire user region at all possible altitudes.

As an example, the radiated region associated with $\varphi_e\,{=}\,28^{\circ}$ and $L_1=25$~m is observed to cover the user region partially over the altitude range of $h\,{\in}\,[21,120]$~m where we have $\varphi_e\,{<}\,\varphi_r$. Indeed, this altitude range is particularly important for the UAV-BS operation, since hovering at lower altitudes ($h\,{<}\,20$~m) is not recommended due to safety issues while higher altitudes ($h\,{>}\,120$~m) is restricted due to regulations of authorities in charge~\cite{FAARule}. When the inner radius $L_1$ is made even larger, the required vertical beamwidth $\varphi_r$ to cover the entire user region becomes relatively small as can be visually inferred from Fig.~\ref{fig:VertAngle}. For example, the specific value $L_1\,{=}\,85$~m might well correspond to a regular football game setting of a stadium, where users are allowed to seat only in tribunes. These intuitive results show that achievable coverage over the user region heavily depends on the interaction among the vertical beamwidth of the UAV-BS beam, user deployment setting, and the operational altitude of UAV-BS, which will be investigated rigorously in the subsequent sections.

\subsection{Beam Scanning over User Region}\label{Sec:Beam_Scanning}
When the physically radiated region by the UAV-BS beam is smaller than the user region, it may matter which portion of the user region should be covered during the DL transmission. By changing the vertical tilting angle of the antenna array, the intersection point of the boresight direction of the UAV beam and the horizontal plane can be moved radially forward (towards outer radius) and backward (towards inner radius), as shown in Fig.~\ref{fig:footprint}. This way, it is possible to change the average path loss and radiated region size, both of which are affecting to the user sum rate. It is therefore of particular interest to search for an optimal intersection point or, equivalently, coverage within the user region for a given vertical beamwidth, $\varphi_{e}$ which is insufficient to cover the entire user region ($\varphi_{e}\,{<}\,\varphi_r$), at a particular UAV altitude.

We assume that the distance to the boresight intersection point from the origin is represented by $D$ (see Fig.~\ref{fig:footprint}). Keeping the radiated region fully inside the user region, we define $D_1$ and $D_2$ to be the two extreme values of $D$ where the inner-most and the outer-most portions of the user region are being covered, respectively, as in Fig.~\ref{fig:footprint}. As a result, $D_1$ corresponds to the radiated region where $\l_1\,{=}\,L_1$ and $\l_2\,{<}\,L_2$, whereas $D_2$ corresponds to the scenario of $\l_1\,{>}\,L_1$ and $\l_2\,{=}\,L_2$. With this, our proposed beam scanning strategy aims to find the optimum boresight intersection point $D^*$, or equivalently, the optimal coverage, such that the NOMA sum rates are maximized at a given altitude $h$. This can be formulated as follows
\begin{align} \label{eq:Opt_D}
D^{*} = \argmax_{D_1 \,{\leq}\, D \,{\leq}\, D_2} R^{\textrm{NOMA}},
\end{align}
where $D_1\,{=}\,h \tan \left(\tan^{{-}1}(L_1/h) \,{+}\, \varphi_e/2 \right)$ and $D_2\,{=}\,h \tan \left(\tan^{{-}1}(L_2/h) \,{-}\, \varphi_e/2 \right)$ via the geometry of Fig.~\ref{fig:footprint}, and $R^{\textrm{NOMA}}$ is the NOMA sum rate. The optimum boresight intersection point $D^*$ yields the optimum downward tilting angle as well  for UAV-BS transmission at the given altitude. Note that we can formulate the optimization problem in \eqref{eq:Opt_D} differently by adding some other constraints (e.g., the existence of both users simultaneously in the radiated region), and even adopt different solution techniques instead of greedy search. Either way, it would still be possible to employ proposed analytical framework developed here while evaluating the respective outage sum rates.

\section{NOMA for UAV-BS Downlink} \label{Sec:NOMA_Transmission}
In this section, we consider NOMA transmission to serve multiple users simultaneously, which are called NOMA users hereafter, using a single DL beam of UAV-BS as sketched in Fig.~\ref{fig:footprint}. Assuming that each user has its own QoS based target rate, we evaluate respective sum rates to investigate conditions to serve each user at least at its target rate.

\subsection{Outage Probabilities and Sum Rates for NOMA}\label{sec:sumrate_noma}
In order to investigate sum rates for NOMA transmission, we first evaluate outage probability of each user individually, which captures the probability of a user being served at a rate less than its target rate. The sum rates are then computed as the weighted sum of target rates, where each target rate is weighted by its non-outage probability, and, hence, are called \textit{outage} sum rates. To this end, we first characterize the effective channel gains for each user, which will then be used to derive outage probabilities and outage sum rates.

We assume that the single UAV-BS may be assigned to either the entire environment where users are distributed, e.g., a stadium, or a part of it, e.g., a sector of a stadium. The AoD $\overline{\theta}$ of the beam $\textbf{b}$ generated by UAV-BS is therefore assumed to take values either from $[0{,}\,2\pi]$, or a subset of it. In addition, the full coverage of the entire environment can be performed by choosing values for $\overline{\theta}$ from its support either sequentially or randomly over time. Without any loss of generality, the effective channel gain of user $k\,{\in}\,\mathcal{N}_\textrm{U}$ for a beamforming direction $\overline{\theta}$ of UAV beam $\textbf{b}= \frac{1}{\sqrt{M}} \left[ 1 \;\; e^{-j2\pi \frac{D}{\zeta}\sin( \overline{\theta})} \; \dots \; e^{-j2\pi \frac{D}{\zeta}\left( M-1\right)\sin (\overline{\theta}) } \right]^{\rm T}$ is given using \eqref{k_UE_original_channel} as follows
\begin{align} \label{eq:Eff_channel_gain_original}
|\textbf{h}_k^{\rm H}\textbf{b}|^2 &= \frac{|\alpha_k|^2 |\textbf{b}^{\rm H}\textbf{a}(\theta_k)|^2}{M  \times \textrm{PL}\left(\sqrt{d_k^2 + h^2}\right)}
= \frac{|\alpha_k|^2 M}{\textrm{PL}\left(\sqrt{d_k^2 + h^2}\right)}
 \left| \frac{ \sin \left( \frac{\pi M(\sin \overline{\theta}-\sin \theta_k)}{2} \right)}{  M\sin \left( \frac{\pi (\sin \overline{\theta}-\sin \theta_k)}{2} \right)}\right|^2 \,
\end{align}
where we assume a critically spaced ULA, i.e. $D = \frac{\zeta}{2}$. Following the convention of~\cite{Ding17PoorRandBeamforming}, we assume small $2\Delta$ while analyzing sum rates, i.e., $2\Delta\,{\rightarrow}\,0$, which results in small angular offset such that $|\overline{\theta}\,{-}\,\theta_k|\,{\rightarrow}\,0$. Choosing the coordinate system appropriately, small angular offset implies small individual angles such that $\sin\overline{\theta}\,{\rightarrow}\,\overline{\theta}$ and $\sin\theta_k\,{\rightarrow}\,\theta_k$, and \eqref{eq:Eff_channel_gain_original} can be approximated as
\begin{align} \label{eq:Eff_channel_gain}
|\textbf{h}_k^{\rm H}\textbf{b}|^2 \approx \frac{|\alpha_k|^2}{M \times\textrm{PL}\left(\sqrt{d_k^2 + h^2}\right)}
 \left| \frac{ \sin \left( \frac{\pi M(\overline{\theta} - \theta_k)}{2} \right)}{  \sin \left( \frac{\pi (\overline{\theta} - \theta_k)}{2} \right)}\right|^2 = \frac{|\alpha_k|^2}{\textrm{PL}\left(\sqrt{d_k^2 + h^2}\right)}  {\rm F}_M(\pi [\overline{\theta} - \theta_k]),
\end{align}
where ${\rm F}_M(\cdot)$ is called Fej\'er kernel. From \eqref{eq:Eff_channel_gain} we observe that $k$-th user effective channel gain depends on the offset of its angle $\theta_k$ from the beamforming direction $\overline{\theta}$, horizontal distance $d_k$ to the UAV-BS, and path gain $\alpha_k$. Note that although $2\Delta\,{\rightarrow}\,0$ is not a necessary condition for our analysis, the effective channel gain in \eqref{eq:Eff_channel_gain} explicitly captures the impact of the user angle $\theta_k$.

NOMA transmission suggests to allocate power to each NOMA user in a way inversely proportional to its channel quality, and therefore requires the ordering of users based on their channel qualities. Deferring the discussion of user ordering strategies to the next section, we assume without any loss of generality that the users in set $\mathcal{N}_{\rm U}$ are already indexed from the worst to the best channel quality under a given criterion. Defining $\beta_k$ to be the power allocation coefficient of $k$-th user, we therefore have $\beta_1\,{\geq}\,\dots\,{\geq}\,\beta_K$ such that $\sum \limits_{k{=}1}^{K} \beta_k^2\,{=}\,1$. The transmitted signal $\textbf{x}$ is then generated by superposition coding as follows
\begin{align} \label{eq:Tx_signal}
\textbf{x} = \sqrt{P_{\rm Tx}}\textbf{b}\sum \limits_{k = 1}^{K} \beta_k s_{k}
\end{align}
where $P_{\rm Tx}$ and $s_{k}$ are the total DL transmit power and $k$-th user's message, respectively. Here $\mathbb{E}\left( \left| s_k \right|^2 \right) = 1$. With \eqref{eq:Tx_signal}, received signal at $k$-th user is given as
\begin{align}\label{eq:k_th_user_Observation}
y_{k}= \textbf{h}_{k}^{\rm H} \textbf{x} +  v_k = \sqrt{P_{\rm Tx}}\textbf{h}_{k}^{\rm H} \textbf{b}\sum \limits_{k = 1}^{K} \beta_k s_{k} + v_k,
\end{align}
where $v_k$ is zero-mean complex Gaussian additive white noise with variance $N_0$.

Adopting successive interference cancellation (SIC) approach, $k$-th user first decodes messages of weaker users (allocated with larger power) in the presence of stronger users' messages (allocated with smaller power), and then subtracts decoded messages from its received signal $y_k$ in \eqref{eq:k_th_user_Observation}. Thus, at $k$-th user, message intended to user $m$  will be decoded with the following SINR
\begin{align} \label{eq:SINR_mk_th_user}
\textrm{SINR}_{m{\rightarrow}k} = \frac{P_{\rm Tx}|\textbf{h}_{k}^{\rm H}\textbf{b}|^2 \beta_{m}^2}{P_{\rm Tx} \sum \limits_{l = m+1}^{K}|\textbf{h}_{k}^{\rm H}\textbf{b}|^2 \beta_{l}^2 + N_0},
\end{align}
where $1 \,{\leq}\, m \,{\leq}\, k-1$. Assuming that all interfering messages of weaker users are decoded accurately, which requires the instantaneous rate associated with decoding any of these weaker users' messages to be larger than the respective target rate of that user, $k$-th user has the following SINR while decoding its own message
\begin{align} \label{eq:SINR_k_th_user}
{\rm SINR}_{k} =  \frac{P_{\rm Tx}|\textbf{h}_{k}^{\rm H}\textbf{b}|^2 \beta_{k}^2}{ \left( 1- \delta_{kK} \right) P_{\rm Tx}  \sum \limits_{l = k+1}^{K}|\textbf{h}_{k}^{\rm H}\textbf{b}|^2 \beta_{l}^2 + N_0}.
\end{align} Here, $\delta_{kK}$ is the Kronecker delta function taking $1$ if $k\,{=}\,K$, and $0$ otherwise. Next, we study how to evaluate outage probabilities and then outage sum rates using SINR terms in \eqref{eq:SINR_mk_th_user} and \eqref{eq:SINR_k_th_user} considering different NOMA formulation criteria.

Defining the instantaneous rates associated with \eqref{eq:SINR_mk_th_user} and \eqref{eq:SINR_k_th_user} to be $R_{m{\rightarrow}k}\,{=}\,\log_2 \left( 1\,{+}\,\textrm{SINR}_{m{\rightarrow}k}\right)$ and $R_{k}\,{=}\,\log_2 \left( 1\,{+}\,\textrm{SINR}_{k}\right)$, respectively, the outage probability of $k$-th NOMA user is given as
\begin{align}
\textrm{P}_{k}^{o} &= 1 - \textrm{P} \left( R_{1{\rightarrow}k} > \overline{R}_1, \dots , R_{k{-}1{\rightarrow}k} > \overline{R}_{k{-}1}, R_{k} > \overline{R}_k | \, \mathcal{S}_K \right) \label{eq:Outage_k_th_user_1}\\
&= 1 - \textrm{P} \left( \textrm{SINR}_{1{\rightarrow}k} > \epsilon_1, \dots , \textrm{SINR}_{k{-}1{\rightarrow}k} > \epsilon_{k{-}1}, \textrm{SINR}_{k} > \epsilon_k | \, \mathcal{S}_K \right) \label{eq:Outage_k_th_user_2},
\end{align}
where $\overline{R}_k$ is the QoS based target rate for $k$-th user and $\epsilon_k\,{=}\,2^{\overline{R}_k}\,{-}\,1$. Note that \eqref{eq:Outage_k_th_user_1} and \eqref{eq:Outage_k_th_user_2} are defined for $\mathcal{S}_K$ describing the given condition on $K$ which might involve either a range of integers, i.e., $\mathcal{S}_K\,{:}\,\left\lbrace K \,|\, j\,{\leq}\, K \,{<}\, i \right\rbrace$, or a unique integer, i.e., $\mathcal{S}_K\,{:}\, \left\lbrace K \,|\, K\,{=}\,i \right\rbrace$, where $i,\, j \in \mathbb{Z}^+$.

When $\mathcal{S}_K$ denotes a unique $K$ value, the outage sum rate is given as
\begin{align} \label{eq:sum_rate_NOMA_singleK}
R^{\textrm{NOMA}} &= \textrm{P} \left( K=1 \right) \left(1 - \tilde{\textrm{P}}_{1}^{o} \right) \overline{R}_1 +
\sum\limits_{n=2}^{\infty} \textrm{P} \left( K=n \right) \left( \sum\limits_{k=1}^{n} \left( 1- \textrm{P}_{k}^{o} \right) \overline{R}_k \right),
\end{align}
where $\tilde{\textrm{P}}_{k}^{o}\,{=}\,\textrm{P}\left(\frac1K \log\left(1{+}P_{\rm Tx}|\textbf{h}_{k}^{\rm H}\textbf{b}|^2/N_0 \right)\,{<}\,\overline{R}_k|\mathcal{S}_K \right)$ is the outage probability of $k$-th user during OMA transmission with the factor $\frac1K$ capturing the loss of degrees-of-freedom (DoF) gain due to OMA. For performance comparison, we consider OMA sum rates which are computed the same way as in \eqref{eq:sum_rate_NOMA_singleK}, except that $\textrm{P}_{k}^{o}$ in the inner summation should be substituted with $\tilde{\textrm{P}}_{k}^{o}$.

Similarly, when we have the set $\mathcal{S}_K\,{:}\,\left\lbrace K \,|\, j\,{\leq}\, K \,{<}\, i \right\rbrace$, the outage sum rate is
\begin{align} \label{eq:sum_rate_NOMA_setK}
R^{\textrm{NOMA}} = \sum \limits_{k = 1}^{K}  (1- \textrm{P}_{k}^{o}) \overline{R}_k,
\end{align}
where $K\,{\in}\,\mathcal{S}_K$. Note that whenever we have $K\,{=}\,1$, single user transmission is employed where the full time-frequency resources and transmit power are allocated to the scheduled user. Note also that OMA sum rates can be readily computed by using \eqref{eq:sum_rate_NOMA_setK} and replacing $\textrm{P}_{k}^{o}$ with $\tilde{\textrm{P}}_{k}^{o}$.

\subsection{Full CSI and Distance Feedback}\label{sec:feedback_noma}
Since NOMA transmitter allocates power to NOMA users based on their channel qualities, it needs to order users according to their effective channel gains which requires feedback of the full CSI of each user. Following the convention of the previous section, this order is given as
\begin{align} \label{Sorted_Ch_gain}
|\textbf{h}_1^{\rm H}\textbf{b}|^2 \leq |\textbf{h}_2^{\rm H}\textbf{b}|^2 \leq \cdots \leq |\textbf{h}_{K}^{\rm H}\textbf{b}|^2.
\end{align}
where the indices of users in set $\mathcal{N}_{\rm U}$ are arranged, without any loss of generality, such that user $k$ has the $k$-th smallest effective channel gain. As a result, $i$-th user has a weaker channel quality than $j$-th user with $i\,{<}\,j$, which meets the power allocation strategy $\beta_i\,{\geq}\,\beta_j$ of Section~\ref{sec:sumrate_noma}.

When the underlying channel experiences rapid fluctuations over time, tracking of CSI becomes computationally inefficient, and frequently sending this information back to the transmitter increases overhead. Note that the horizontal distance is one factor directly affecting the channel quality as shown in \eqref{eq:Eff_channel_gain}, and does not vary fast as compared to the effective channel gain. Thus, we consider to use distance information as a practical alternative of full CSI for ordering users. Since the effective channel gain is inversely proportional to horizontal distance as in \eqref{eq:Eff_channel_gain}, the following order is assumed for this \textit{limited feedback} scheme
\begin{align}\label{eq:Sorted_Distances}
d_1 \leq d_2 \leq\dots \leq d_K.
\end{align}
With this order, $i$-th user has still weaker channel quality than $j$-th user, but this time with $i\,{>}\,j$, and we have still the same power allocation $\beta_i\,{\geq}\,\beta_j$. In addition, formulation of outage probabilities and sum rates of Section~\ref{sec:sumrate_noma} applies to the distance feedback scheme, as well. However, the indices in set $\mathcal{N}_{\rm U}$ are now representing users ordered from the best to the worst channel quality such that user $k$ is assumed to have the $k$-th best channel quality.

\subsection {Multiple Access for Partial Coverage} \label{sec:noma_events}
When the required vertical beamwidth is greater than the available UAV-BS vertical beamwidth, i.e., $\varphi_r\,{>}\,\varphi_e$, the user region is covered partially, and we explore the optimal area for the radiated region within the user region through beam scanning approach, as described in Section~\ref{Sec:Beam_Scanning}. Note that it might not be possible to find desired NOMA users in the radiated region when the user region is partially covered. In the following, we will elaborate all possible conditions for the presence of NOMA users within the radiated region considering $i$-th and $j$-th users only, though results can be generalized to multiple NOMA users as well.

Denoting the set of users inside the radiated region by $\mathcal{N}_{\rm U}^{D}$, we identify four possible mutually exclusive events for the presence of the $i$-th and $j$-th users within the radiated region as follows
\begin{itemize}
\item \textit{Event $1$} ($\textrm{E}_1$): Both users are outside the radiated region $\left( i,j \notin \mathcal{N}_{\rm U}^{D} \right)$,
\item \textit{Event $2$} ($\textrm{E}_2$): Only $i$-th user is within the radiated region $\left( i \in \mathcal{N}_{\rm U}^{D}, j \notin \mathcal{N}_{\rm U}^{D} \right)$,
\item \textit{Event $3$} ($\textrm{E}_3$): Only $j$-th user is within the radiated region $\left( j \in \mathcal{N}_{\rm U}^{D}, i\notin \mathcal{N}_{\rm U}^{D} \right)$,
\item \textit{Event $4$} ($\textrm{E}_4$): Both users are within the radiated region $\left( i, j \in \mathcal{N}_{\rm U}^{D} \right)$.
\end{itemize}

We modify the NOMA transmission strategy based on these four possible events as follows. The original NOMA is applicable only for $\textrm{E}_4$ since it is the only case having both users within the radiated region. When $\textrm{E}_2$ or $\textrm{E}_3$ occurs, $i$-th or $j$-th user, respectively, will be served all the time with full transmit power, which is therefore called single user transmission, and is common to OMA  with the same outage probability. And, finally, no DL transmission will take place for $\textrm{E}_1$ as both users will be in outage. The overall transmission strategy is referred to as \textit{hybrid NOMA}, which highlights the fact that UAV-BS employs NOMA whenever both users are available, and switches to single user transmission if a single NOMA user is present in the radiated region.

\section{NOMA Outage Sum Rate with Distance Feedback} \label{Sec:NOMARates_Distance_FB}
In this section, we will provide analytical expressions for the outage sum rate of NOMA transmission strategy described in Section~\ref{sec:noma_events}, when distance feedback is employed. We consider to serve $i$-th and $j$-th users only with $i\,{>}\,j$, which are designated as the weaker and stronger users, respectively, and results can be generalized to multiple NOMA users, as well. We first formulate the outage sum rate expression based on events of Section~\ref{sec:noma_events} and respective outage probabilities, and then derive all these probability expressions assuming number of users to satisfy $j\,{\leq}\, K \,{<}\, i$ and $K \,{\geq}\, i$, in sequence. Finally, we provide an asymptotic analysis of outage probabilities to obtain further insight.

\subsection{Outage Sum Rate Formulation}
We first formulate outage sum rate expression for NOMA transmission based on events captured in Section~\ref{sec:noma_events} and respective outage probabilities, for $i$-th and $j$-th users with $i\,{>}\,j$. Note that, there should be at least $j$ users in the user region to start transmission, and $K\,{\geq}\,j$. We will evaluate the overall performance by individually considering the number of users $K$ to be in sets $\mathcal{S}_{K_{1}}{:}\left\lbrace K \,|\, j\,{<}\, K \,{\leq}\, i \right\rbrace$ and $\mathcal{S}_{K_2}{:}\left\lbrace K  \,|\, K \,{\geq}\, i \right\rbrace$ separately, and then combine them statistically to yield the desired result for $K\,{\geq}\,j$.

\begin{table}[!h]
\centering
\caption{Possible NOMA Events with Nonzero Probability.}
\label{tab:noma_events}
{\normalsize
\begin{tabular}{c|c|c|}
\cline{2-3}
	& $\mathcal{S}_{K_{1}} \,{:}\, j\,{\leq}\, K \,{<}\, i$ & $\mathcal{S}_{K_{2}} \,{:}\, K \,{\geq}\, i$	\\
\hhline{-==}
\multicolumn{1}{|c|}{$\varphi_e\,{<}\,\varphi_r$} & $\textrm{E}_1$, $\textrm{E}_3$	& $\textrm{E}_1$, $\textrm{E}_2$, $\textrm{E}_3$, $\textrm{E}_4$ \\
\hline
\multicolumn{1}{|c|}{$\varphi_e\,{\geq}\,\varphi_r$} 	& $\textrm{E}_3$					& $\textrm{E}_4$ \\
\hline
\end{tabular}}
\end{table}
In Table~\ref{tab:noma_events}, we list all possible NOMA events with nonzero probability based on the number of users, and the status of user region coverage. As an example, we have only $j$-th user for $j\,{\leq}\, K \,{<}\, i$, which might or might not be present in the radiated region captured by the events $\textrm{E}_3$ and $\textrm{E}_1$, respectively, when the user region is partially covered with $\varphi_e\,{<}\,\varphi_r$. Note that, whenever the user region is fully covered  with $\varphi_e\,{\geq}\,\varphi_r$, $\textrm{E}_3$ and $\textrm{E}_4$ are the only possible events for $\mathcal{S}_{K_{1}}$ and $\mathcal{S}_{K_{2}}$, respectively, hence respective event probabilities are $1$. We therefore consider the derivation of event probabilities only when the user region is partially covered.

The general expression for outage sum rates is therefore given as
\begin{align} \label{eq:outage_sumrate}
R^{\textrm{NOMA}} &= {\rm P} \left\lbrace \mathcal{S}_{K_{1}} \right\rbrace \left[ {\rm P} \{ {\rm E}_3 \} \left( 1 - {\rm P}_{j|\mathcal{S}_{K_1}}^{\rm o,\,3} \right) \overline{R}_j \right] +  {\rm P} \{\mathcal{S}_{K_{2}}\} \left[ \textrm{P} \{ {\rm E}_2 \} \left( 1 - {\rm P}_{i|\mathcal{S}_{K_2}}^{\rm o,\,2} \right) \overline{R}_i \right.
\nonumber \\
&  \quad + \left.\textrm{P} \{{\rm E}_3\} \left( 1 - {\rm P}_{j|\mathcal{S}_{K_2}}^{\rm o,\,3} \right) \overline{R}_j
+ \textrm{P} \{{\rm E}_4\} \left( (1 - {\rm P}_{i|\mathcal{S}_{K_2}}^{\rm o,\, 4} ) \overline{R}_i + \left(1 - {\rm P}_{j|\mathcal{S}_{K_2}}^{\rm o, \, 4} \right) \overline{R}_j \right) \right],
\end{align}
where ${\rm P}\{ \mathcal{S}_{K_1} \}$ and ${\rm P} \{ \mathcal{S}_{K_2}\}$ represent the probability of having $K$ users in the user region given by $\mathcal{S}_{K_1}$ and $\mathcal{S}_{K_2}$, respectively, $\textrm{P} \{{\rm E}_n\}$ is the probability of event~$n$, and ${\rm P}_{k|\mathcal{S}_{K_l}}^{\rm o,\,n}$ is the \textit{conditional} outage probability of $k$-th user for a given event $n$ and set $\mathcal{S}_{K_l}$, with $k\,{\in}\,\{i,j\}$, $n\,{\in}\,\{2,3,4\}$, and $l\,{\in}\,\{1,2\}$. Note that, \eqref{eq:outage_sumrate} is applicable to both full and partial coverage of user region through suitable event probabilities such that $\textrm{P} \{{\rm E}_2\}\,{=}\,\textrm{P} \{{\rm E}_3\}\,{=}\,0$ and $\textrm{P} \{{\rm E}_3\}\,{=}\,\textrm{P} \{{\rm E}_4\}\,{=}\,1$ under full coverage. Since we consider a range of $K$ values, \eqref{eq:outage_sumrate} can be rearranged to yield \eqref{eq:sum_rate_NOMA_setK}, where the overall \textit{unconditional} outage probabilities of $i$-th and $j$-th users are given as
\begin{align}
{\rm P}_j^{\rm o} &= 1-\left[{\rm P} \left\lbrace \mathcal{S}_{K_1} \right\rbrace  {\rm P} \{ {\rm E}_3 \} \left( 1 - {\rm P}_{j|\mathcal{S}_{K_1}}^{\rm o,\,3} \right) \right. \nonumber \\
& \quad \qquad \left. + \; {\rm P} \{\mathcal{S}_{K_2}\} \left\lbrace \textrm{P} \{{\rm E}_3\} \left( 1 - {\rm P}_{j|\mathcal{S}_{K_2}}^{\rm o,\,3} \right) + \textrm{P} \{{\rm E}_4\} \left(1 - {\rm P}_{j|\mathcal{S}_{K_2}}^{\rm o, \, 4} \right) \right\rbrace \right] , \label{eq:outage_final_j} \\
{\rm P}_i^{\rm o} &= 1 -  {\rm P} \{\mathcal{S}_{K_2}\} \left[ \textrm{P} \{ {\rm E}_2 \} \left( 1 - {\rm P}_{i|\mathcal{S}_{K_2}}^{\rm o,\,2} \right) + \textrm{P} \{{\rm E}_4\} \left( 1 - {\rm P}_{i|\mathcal{S}_{K_2}}^{\rm o,\, 4} \right) \right] \label{eq:outage_final_i},
\end{align} and, outage sum rates of \eqref{eq:sum_rate_NOMA_setK} and \eqref{eq:outage_sumrate} can be calculated as $R^{\textrm{NOMA}}\,{=}\, {\rm P}_i^{\rm o} \overline{R}_i \,{+}\, {\rm P}_j^{\rm o} \overline{R}_j$. In the subsequent sections, we derive analytical expressions for event probabilities $\textrm{P} \{{\rm E}_n\}$ under partial coverage of user region, and conditional outage probabilities ${\rm P}_{k|\mathcal{S}_{K_l}}^{\rm o,\,n}$ given in \eqref{eq:outage_sumrate} for the sets $\mathcal{S}_{K_1}$ and $\mathcal{S}_{K_2}$, in sequence, assuming both coverage status.

\subsection{Event and Conditional Outage Probabilities for $\mathcal{S}_{K_1}$} \label{sec:Events_for_j_less_K_less_i}
In order to analytically evaluate event and conditional outage probabilities in \eqref{eq:outage_sumrate} for $\mathcal{S}_{K_1}$, we first consider the marginal PDF of the $k$-th user distance conditioned on $\mathcal{S}_{K_1}$.
\begin{theorem}\label{the:marginal_pdf}
Assuming that the number of users $K$ is chosen from the set $\mathcal{S}_{K_1}$ such that $j \,{\leq}\, K \,{<}\, i$, the marginal PDF of the $k$-th user distance $d_k$ is given as
\begin{align} \label{eq:PDF_j_less_K_less_i}
f_{d_k|\mathcal{S}_{K_1}}(r_k) = \frac{ 2 \Delta \lambda r_k}{\mathcal{C}}e^{-\Delta(L_2^2 - L_1^2)\lambda}\frac{\left[\Delta(r_k^2 - L_1^2)\lambda \right]^{(k-1)}}{(k-1)!} \left( \sum \limits_{l=0}^{i-k-1}\frac{\left[\Delta(L_2^2 - r_k^2)\lambda \right]^{l}}{l!} \right)
\end{align}
where $\mathcal{C}\,{=}\,\sum\limits_{l=j}^{i{-}1} \frac{e^{{-}\Delta(L_2^2\,{-}\,L_1^2)\lambda} \left[\Delta(L_2^2 \,{-}\, L_1^2)\lambda \right]^l}{l!}$.
\end{theorem}
\begin{IEEEproof}
See Appendix~\ref{app:PDF_j_less_K_less_i}.
\end{IEEEproof}

Since this particular case assumes the presence of only $j$-th user and no $i$-th user at all, the possible events are ${\rm E}_1$ and ${\rm E}_3$, as shown in Table~\ref{tab:noma_events}. As there is no transmission for ${\rm E}_1$, we focus on ${\rm E}_3$ in this particular case. For partially covered user region with $\varphi_e \,{<}\, \varphi_r$, ${\rm E}_3$ happens when $j$-th user lies in the radiated region, and $d_j$ is lying in $l_1 \,{\leq}\, d_j \,{\leq}\, l_2$ (see Fig.~\ref{fig:footprint}). Employing \eqref{eq:PDF_j_less_K_less_i}, desired event probability is calculated as
\begin{align} \label{eq:prob_e3_sk1}
{\rm P} \{ {\rm E}_3 \} &= {\rm P} \left\lbrace l_1 \leq d_j \leq l_2 \,{|}\,\mathcal{S}_{K_1} \right\rbrace
= \int \limits_{l_1}^{l_2}f_{d_j|\mathcal{S}_{K_1}}(r) \dd r.
\end{align}

Employing \eqref{eq:SINR_k_th_user} and \eqref{eq:Outage_k_th_user_2}, conditional outage probability for this particular case is given as
\begin{align}
{\rm P}_{j|\mathcal{S}_{K_1}}^{o,\, 3} &= {\rm P} \left\lbrace  \left| \textbf{h}_j^{\rm H} \textbf{b} \right|^2  < \eta_j \,{\big|}\, {\rm E}_3 \right\rbrace
= \frac{1}{{\rm P} \{ {\rm E}_3 \}} {\rm P} \left\lbrace  \left| \textbf{h}_j^{\rm H} \textbf{b} \right|^2  < \eta_j , l_{\rm min} \leq d_j \leq l_{\rm max} \,{|}\,\mathcal{S}_{K_1} \right\rbrace , \label{eq:cond_outage_j_sk1_0} \\
&= \frac{1}{{\rm P} \{{\rm E}_3\}} \int \limits_{\mathcal{D}_{\theta}} {\rm P} \left\lbrace  \left| \textbf{h}_j^{\rm H} \textbf{b} \right|^2  < \eta_j , l_{\rm min} \leq d_j \leq l_{\rm max} \,{\big|}\, x_j, \mathcal{S}_{K_1} \right\rbrace
 f_{x_j} (x) \dd x , \label{eq:cond_outage_j_sk1_1}
\end{align}
where $f_{x_j}(x)$ is the PDF of $j$-th user location $x_j$, $\mathcal{D}_{\theta}$ denotes the radiated region, and $\eta_j\,{=}\,\frac{\epsilon_j}{P_{\rm Tx}/N_0}$. Note that both full and partial coverage of the user region is considered in \eqref{eq:cond_outage_j_sk1_0} by choosing $d_j$ interval with the limits $l_{\rm min}\,{=}\,\max(L_1,l_1)$ and $ l_{\rm max}\,{=}\,\min(l_2,L_2)$. Since $x_j$ is fully specified by distance $d_j$ and angle $\theta_j$, which are independent of each other with angle $\theta_j$ being uniformly distributed within $\left[\overline{\theta}\,{-}\,\Delta{,}\,\overline{\theta}\,{+}\,\Delta \right]$ due to the specific geometry of the radiated region, \eqref{eq:cond_outage_j_sk1_1} can be represented as
\begin{align}
{\rm P}_{j|\mathcal{S}_{K_1}}^{o,\, 3} &= \frac{1}{{\rm P} \{{\rm E}_3\}} \int\limits_{\overline{\theta}{-}\Delta}^{\overline{\theta}{+}\Delta} \int\limits_{l_1}^{l_2} {\rm P} \left\lbrace \left| \textbf{h}_j^{\rm H}\textbf{b} \right|^2  < \eta_j \,{\big|}\, d_j\,{=}\,r,\,\theta_j\,{=}\,\theta \right\rbrace \frac{f_{d_j|\mathcal{S}_{K_1}}(r) }{2\Delta} \dd r \dd \theta , \label{eq:cond_outage_j_sk1_2} \\
&=\frac{1}{{\rm P} \{{\rm E}_3\}} \int\limits_{\overline{\theta}{-}\Delta}^{\overline{\theta}{+}\Delta} \int\limits_{l_1}^{l_2}  \left( 1 - \exp \left\lbrace \frac{-\eta_j{\textrm{PL} \left(\sqrt{r^2 + h^2}\right)}}{ {{\rm F}_M(\pi[\overline{\theta}\,{-}\,\theta])}} \right\rbrace  \right) \frac{f_{d_j|\mathcal{S}_{K_1}}(r) }{2\Delta} \dd r \dd \theta , \label{eq:cond_outage_j_sk1_3}
\end{align}
by employing the distribution of $\left| \textbf{h}_j^{\rm H}\textbf{b} \right|^2$ which is exponential for a given location since path gain $\alpha_k$ is complex Gaussian.

\subsection{Event and Conditional Outage Probabilities for $\mathcal{S}_{K_2}$} \label{sec:Events_for_K_greater_i}
We now consider event and conditional outage probabilities for $\mathcal{S}_{K_2}$, where we assume the presence of both $i$-th and $j$-th users. We therefore need the joint PDF of user distances $d_i$ and $d_j$, which is given in the following theorem.
\begin{theorem}\label{the:joint_pdf}
Assuming that the number of users $K$ is chosen from the set $\mathcal{S}_{K_2}$ such that $K \,{\geq}\, i$, the joint PDF of $i$-th and $j$-th user distances $d_i$ and $d_j$, respectively, with $d_j \,{\leq}\, d_i$, is given as
 \begin{align} \label{eq:JPDF}
f_{d_j,\,d_i|\mathcal{S}_{K_2}} (r_k,r_i)= \frac{\left( 2 \Delta \lambda\right)^2}{\mathcal{C}}r_k r_i e^{-\Delta(r_i^2 - L_1^2)\lambda} \frac{\left[\Delta(r_k^2 - L_1^2)\lambda \right]^{(j-1)}}{(j-1)!} \frac{\left[\Delta(r_i^2 - r_k^2)\lambda \right]^{(i-j-1)}}{(i-j-1)!}
\end{align} where $\mathcal{C} = 1 - \sum \limits_{l = 0}^{i-1} \frac{e^{-\Delta(L_2^2 - L_1^2)\lambda} \left[\Delta(L_2^2 - L_1^2)\lambda \right]^l}{l!}$.
\end{theorem}
\begin{IEEEproof}
See Appendix~\ref{app:JPDF_K_greater_i}.
\end{IEEEproof}

For this particular case, we derive probabilities of ${\rm E}_2$, ${\rm E}_3$, and ${\rm E}_4$ when the user region is partially covered with $\varphi_e \,{<}\, \varphi_r$, as shown in Table~\ref{tab:noma_events}. We first consider ${\rm E}_2$ where we have only $i$-th user in the radiated region such that $l_1 \,{\leq}\, d_i \,{\leq}\, l_2$ (see Fig.~\ref{fig:footprint}), and $j$-th user should be outside the radiated region. Given $d_j \,{\leq}\, d_i$, which comes from user ordering of Section~\ref{sec:feedback_noma}, the only possible $d_j$ interval is $L_1 \,{\leq}\, d_j \,{\leq}\, l_1$. The respective event probability is therefore given as
\begin{align} \label{eq:prob_e2_sk2}
{\rm P} \{ {\rm E}_2 \} &= {\rm P} \left\lbrace L_1 \,{\leq}\, d_j \,{\leq}\, l_1, \, l_1 \,{\leq}\, d_i \,{\leq}\, l_2 , \, d_j \,{\leq}\, d_i {|}\, \mathcal{S}_{K_2} \right\rbrace =
\int\limits_{l_1}^{l_2} \int\limits_{L_1}^{l_1} f_{d_j,\,d_i|\mathcal{S}_{K_2}} (r,\ell) \dd r \dd \ell .
\end{align}
Similarly, any event ${\rm E}_n$ can be represented as $\left\lbrace a_n \,{\leq}\, d_j \,{\leq}\, b_n, \, u_n \,{\leq}\, d_i \,{\leq}\, v_n {|}\, \mathcal{S}_{K_2} \right\rbrace$, such that
\begin{align} \label{eq:prob_en_sk2}
{\rm P} \{ {\rm E}_n \} &= \int\limits_{u_n}^{v_n} \int\limits_{a_n}^{b_n} f_{d_j,\,d_i|\mathcal{S}_{K_2}} (r,\ell) \dd r \dd \ell ,
\end{align}
where $n\,{\in}\,\{2,3,4\}$, and the integral limits are given as
\begin{align}\label{eq:interval_limits}
\left( a_n, b_n, u_n, v_n \right) = \begin{cases}
(L_1,l_1,l_1,l_2)	& \text{if } n\,{=}\,2 \\
(l_1,l_2,l_2,L_2)	& \text{if } n\,{=}\,3 \\
(l_{\rm min},d_i,l_{\rm min},l_{\rm max})	& \text{if } n\,{=}\,4
\end{cases},
\end{align}
where $l_{\rm min}\,{=}\,\max(L_1,l_1)$ and $ l_{\rm max}\,{=}\,\min(l_2,L_2)$ make sure the validity of \eqref{eq:interval_limits} for full coverage of the user region, which will be employed during outage computation in the sequel.

Following the strategy of \eqref{eq:cond_outage_j_sk1_0}, conditional outage probabilities ${\rm P}_{i|\mathcal{S}_{K_2}}^{\rm o,\,2}$, ${\rm P}_{i|\mathcal{S}_{K_2}}^{\rm o,\,4}$, ${\rm P}_{j|\mathcal{S}_{K_2}}^{\rm o,\,3}$, and ${\rm P}_{j|\mathcal{S}_{K_2}}^{\rm o,\,4}$ of \eqref{eq:outage_sumrate} can therefore be expressed as
\begin{align}
{\rm P}_{k|\mathcal{S}_{K_2}}^{{\rm o},\, n} &= {\rm P} \left\lbrace  \left| \textbf{h}_k^{\rm H} \textbf{b} \right|^2  < \eta_k^{(n)} \,{\big|}\, {\rm E}_n \right\rbrace
= \frac{1}{{\rm P} \{ {\rm E}_n \}} {\rm P} \left\lbrace  \left| \textbf{h}_k^{\rm H} \textbf{b} \right|^2  < \eta_k^{(n)} ,a_n \,{\leq}\, d_j \,{\leq}\, b_n, \, u_n \,{\leq}\, d_i \,{\leq}\, v_n {|}\, \mathcal{S}_{K_2} \right\rbrace \nonumber \\
&= \frac{1}{{\rm P} \{{\rm E}_n\}} \int\limits_{\mathcal{D}_{\theta}} {\rm P} \left\lbrace  \left| \textbf{h}_k^{\rm H} \textbf{b} \right|^2  < \eta_k^{(n)} , a_n \,{\leq}\, d_j \,{\leq}\, b_n, \, u_n \,{\leq}\, d_i \,{\leq}\, v_n \,{\big|}\, x_k, \mathcal{S}_{K_2} \right\rbrace
 f_{x_k|\mathcal{S}_{K_2}} (x) \dd x , \nonumber
\end{align}
where $\eta_k^{(n)}$ is given for $k\,{\in}\,\{i,j\}$, $n\,{\in}\,\{2,3,4\}$ as $\eta_i^{(2)} \,{=}\, \frac{\epsilon_i}{P_{\rm Tx}/N_0}$, $\eta_j^{(3)}\,{=}\,\frac{\epsilon_j}{P_{\rm Tx}/N_0}$, $\eta_i^{(4)}\,{=}\,\frac{\epsilon_i/(P_{\rm Tx}/N_0)}{\beta_i^2{-}\beta_j^2\epsilon_i}$ and $\eta_j^{(4)}\,{=}\,\max\left\lbrace \frac{\epsilon_i/(P_{\rm Tx}/N_0)}{\beta_i^2{-}\beta_j^2\epsilon_i} ,\, \frac{\epsilon_j}{(P_{\rm Tx}/N_0)\beta_j^2} \right\rbrace$~\cite{Ding17PoorRandBeamforming}. Considering all possible $d_i$ and $d_j$ values, the desired conditional outage probabilities can be computed as follows
\begin{align}
{\rm P}_{k|\mathcal{S}_{K_2}}^{{\rm o},\, n} &= \frac{1}{{\rm P} \{{\rm E}_n\}} \int\limits_{\overline{\theta}{-}\Delta}^{\overline{\theta}{+}\Delta} \int\limits_{u_n}^{v_n} \int\limits_{a_n}^{b_n} {\rm P} \left\lbrace \left| \textbf{h}_k^{\rm H}\textbf{b} \right|^2  < \eta_k^{(n)} \,{\big|}\, d_j\,{=}\,r,\,d_i\,{=}\,\ell,\,\theta_k\,{=}\,\theta \right\rbrace \frac{f_{d_j,d_i|\mathcal{S}_{K_2}}(r,\ell)}{2\Delta} \dd r \dd \ell \dd \theta , \nonumber\\
&= \frac{1}{{\rm P} \{{\rm E}_n\}} \int\limits_{\overline{\theta}{-}\Delta}^{\overline{\theta}{+}\Delta} \int\limits_{u_n}^{v_n} \int\limits_{a_n}^{b_n} \left( 1 \,{-}\, \exp \left\lbrace \frac{-\eta_k^{(n)}{\textrm{PL}\left(\sqrt{\delta_{ki}\ell^2 \,{+}\, \delta_{kj}r^2 \,{+}\, h^2}\right)}}{ {{\rm F}_M(\pi[\overline{\theta}\,{-}\,\theta])}} \right\rbrace  \right) \frac{f_{d_j,d_i|\mathcal{S}_{K_2}}(r,\ell)}{2\Delta} \dd r \dd \ell \dd \theta . \label{eq:cond_outage_kn_sk2_3}
\end{align} Here, $\delta_{ki}$ and $\delta_{kj}$ take $1$ if $k\,{=}\,i$ and $k\,{=}\,j$, respectively, and $0$ otherwise.

\subsection{Asymptotic Analysis of the Outage Probabilities}
In this section, we analyze the asymptotic behavior of the outage probabilities derived in \eqref{eq:cond_outage_j_sk1_3} and \eqref{eq:cond_outage_kn_sk2_3}, which are presented by the following theorem.
\begin{theorem} Assuming 1) $2\Delta \rightarrow 0$, and 2) high SNR $\left( P_{\rm Tx}/N_0\rightarrow \infty \right)$, the approximated conditional outage probability expression for the asymptotic analysis of \eqref{eq:cond_outage_j_sk1_3} is
\begin{align} \label{eq:Asy_outage_Sk1}
{\rm P}_{j|\mathcal{S}_{K_1}}^{o,\, 3} \approx \frac{1}{{\rm P} \{{\rm E}_3\}}\left(  1+\frac{\pi^2 M^2 \Delta^2}{36} \right)\Psi_j \eta_j,
\end{align}
where $ \Psi_j = \int_{l_1}^{l_2}  \frac{{\textrm{PL} \left(\sqrt{r^2 + h^2}\right)} }{M} f_{d_j|\mathcal{S}_{K_1}}(r) \dd r$. Likewise, the asymptotic equivalent of  \eqref{eq:cond_outage_kn_sk2_3} is
\begin{align} \label{eq:Asy_outage_Sk2}
{\rm P}_{k|\mathcal{S}_{K_2}}^{{\rm o},\, n} \approx \frac{1}{{\rm P} \{{\rm E}_n\}}\left(  1+\frac{\pi^2 M^2 \Delta^2}{36} \right)\Psi_{ij} \eta_k^{(n)},
\end{align}
where $k\,{\in}\,\{i,j\}$ and $ \Psi_{ij} = \int_{u_n}^{v_n} \int_{a_n}^{b_n}  \frac{\eta_k^{(n)}}{M} {\textrm{PL}\left(\sqrt{\delta_{ki}\ell^2 \,{+}\, \delta_{kj}r^2 \,{+}\, h^2}\right)} f_{d_j,d_i|\mathcal{S}_{K_2}}(r,\ell) \dd r \dd \ell
$.
\end{theorem}
\begin{IEEEproof}
See Appendix~\ref{app:Asymp_analysis_outage_j_i}.
\end{IEEEproof} Note here that in \eqref{eq:Asy_outage_Sk1}, $\eta_j\,{=}\,\frac{\epsilon_j}{P_{\rm Tx}/N_0}$ and hence diversity gain is $1$. Also note that the diversity gain in \eqref{eq:Asy_outage_Sk2} is $1$ for both $i$-th and $j$-th users since $\eta_k^{(n)}\,{\propto}\,\frac{1}{P_{\rm Tx}/N_0}$.
Hence, this asymptotic analysis verifies that \textit{incorporating the $i$-th (weak) user into the limited feedback based NOMA transmission does not alter the achievable diversity gain of the $j$-th (strong) user. In addition, the diversity gain is independent of the user index}.

\section{Numerical Results} \label{sec:Numerical_results}
In this section, we study in detail the achievable outage sum rates for NOMA and OMA transmissions. By making use of the derived analytical expressions in Section~\ref{Sec:NOMARates_Distance_FB} and through extensive computer simulations, we investigate optimal altitudes for UAV operation to maximize sum rates for the scenario discussed in Section~\ref{sec:Sys_Model}. We consider two path-loss models in our analysis: 1) distance dependent PL model given as $ \textrm{PL}(\sqrt{d^2 + h^2}) \,{=}\, 1 {+} (\sqrt{d^2 + h^2})^{\gamma}$~\cite{Ding17PoorRandBeamforming}, where $\gamma$ is the path-loss exponent and $d$ is the horizontal distance to UAV-BS, and 2) \emph{close-in} (CI) free-space reference distance model for urban micro (UMi) mmWave environment given as $\textrm{PL}((\sqrt{d^2 + h^2}),\,f_{\rm c} ) = 32.4 \,{+}\, 21\log_{10}(\sqrt{d^2 + h^2}) \,{+}\, 20\log_{10}(f_{\rm c})$ \cite{CIwhite}, where $f_{\rm c}$ represents the operating mmWave frequency. The simulation parameters are summarized in Table~\ref{tab:SimParameters}.

\begin{table}
\renewcommand{\arraystretch}{0.9}
\centering
\caption{Simulation Parameters}
\begin{tabular}{ | c | c | }
\hline
\textbf{Parameter}          				& \textbf{Value}  \\
\hline\hline
User distribution          					& Uniform 			\\ \hline
Outer radius, $L_{2}$                   	& $100$~m  			\\ \hline
Inner radius, $L_{1}$                   	& $25$~m  			\\ \hline
Horizontal angular width, $2 \Delta$        & $0.5^{\circ}$ 	\\ \hline
Vertical beamwidth, $\varphi_e$           	& $28^{\circ}$ 		\\ \hline
HPPP density, $\lambda$                 	& $1$ 				\\ \hline
Number of BS antennas, $M$               	& $10$    			\\ \hline
Noise, $N_0$                     	    	& $-35$~dBm  		\\ \hline
Path-loss exponent, $\gamma$      			& $2$ 				\\ \hline
$j$th user target rate, $\overline{R}_j$ 	& $6$~BPCU 			\\ \hline
$i$th user target rate, $\overline{R}_j$ 	& $0.5$~BPCU 		\\ \hline
$j$th user power allocation, $\beta_j^2$ 	& $0.25$ 			\\ \hline
$i$th user power allocation, $\beta_i^2$ 	& $0.75$ 			\\ \hline
UAV-BS operation altitude, $h$             	& $10$~m - $150$~m 	\\ \hline
mmWave operating frequency, $f_{\rm c}$		& $30$~GHz	\\ \hline
\end{tabular}
\label{tab:SimParameters}
\end{table}

\subsection{Performance of Distance Feedback: NOMA vs. OMA}\label{sec:numres:noma_distance}
\begin{figure}[!t]
\vspace{-2em}
\centering
\includegraphics[width=0.60\textwidth]{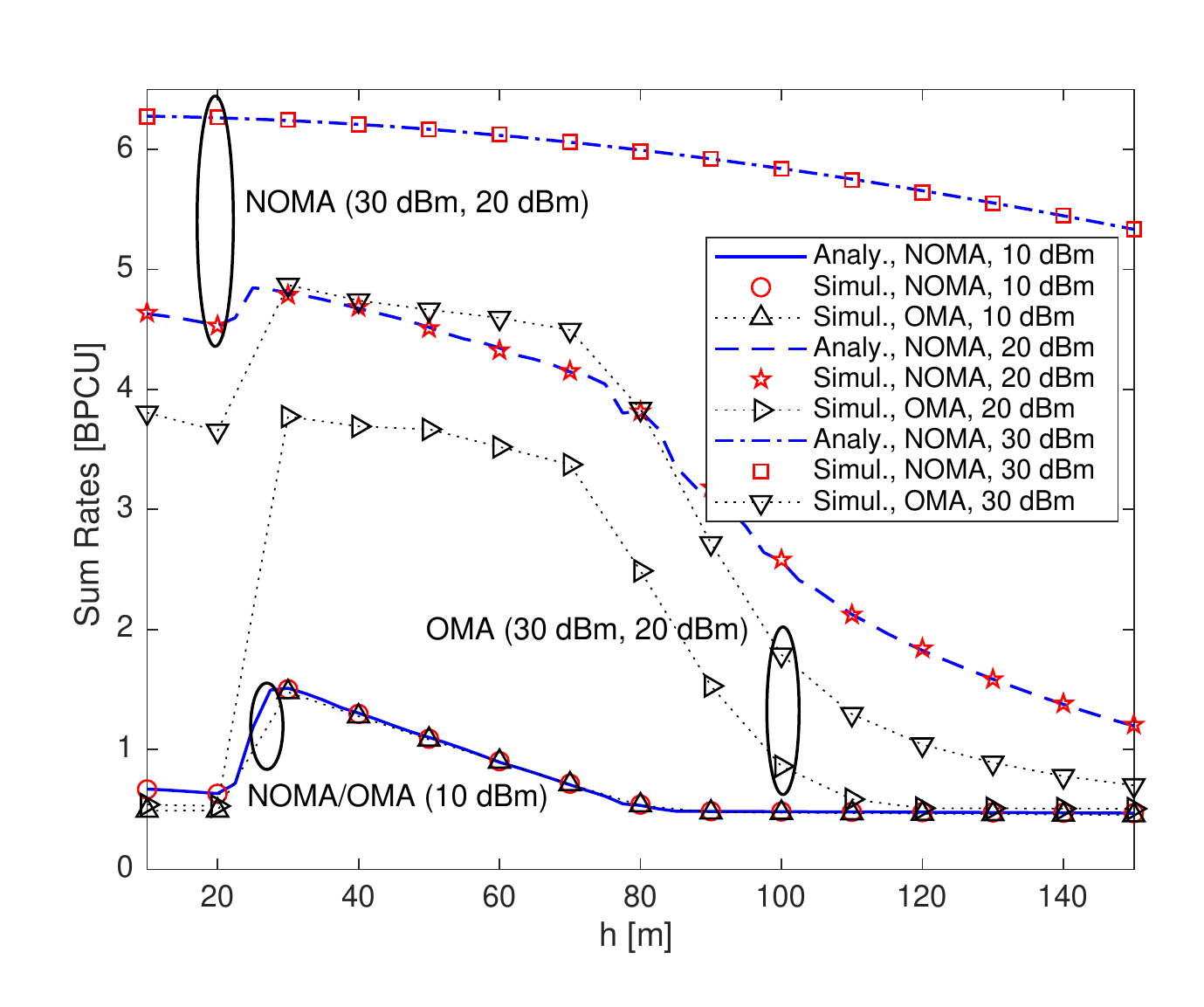}
\caption{Sum rates of OMA and NOMA with distance feedback where $i\,{=}\,30$, $j\,{=}\,20$.}
\label{fig:sumrate_distance_j20_i30}
\end{figure}

In Fig.~\ref{fig:sumrate_distance_j20_i30}, we present sum rate performance of OMA and NOMA for distance feedback scheme along with varying altitude considering $i\,{=}\,30$ and $j\,{=}\,20$ user pair. As can be observed, analytical sum rate results for NOMA perfectly match with the simulation based results. We observe that OMA and NOMA sum rate performance are very similar for $P_{\rm Tx}\,{=}\,10$~dBm, which is also observed from respective outage results of $i$-th and $j$-th users in Fig.~\ref{fig:outage_distance_j20_i30}. On the other hand, when $P_{\rm Tx}\,{=}\,30$~dBm, sum rates of NOMA become significantly better than that of OMA along with relatively lower outage probabilities of both NOMA users as captured in Fig.~\ref{fig:outage_distance_j20_i30}.

\begin{figure}[!t]
\vspace{-2em}
\centering
\hspace*{-0.3in}
\subfloat[$i$-th user with $i\,{=}\,30$.]{\includegraphics[width=0.55\textwidth]{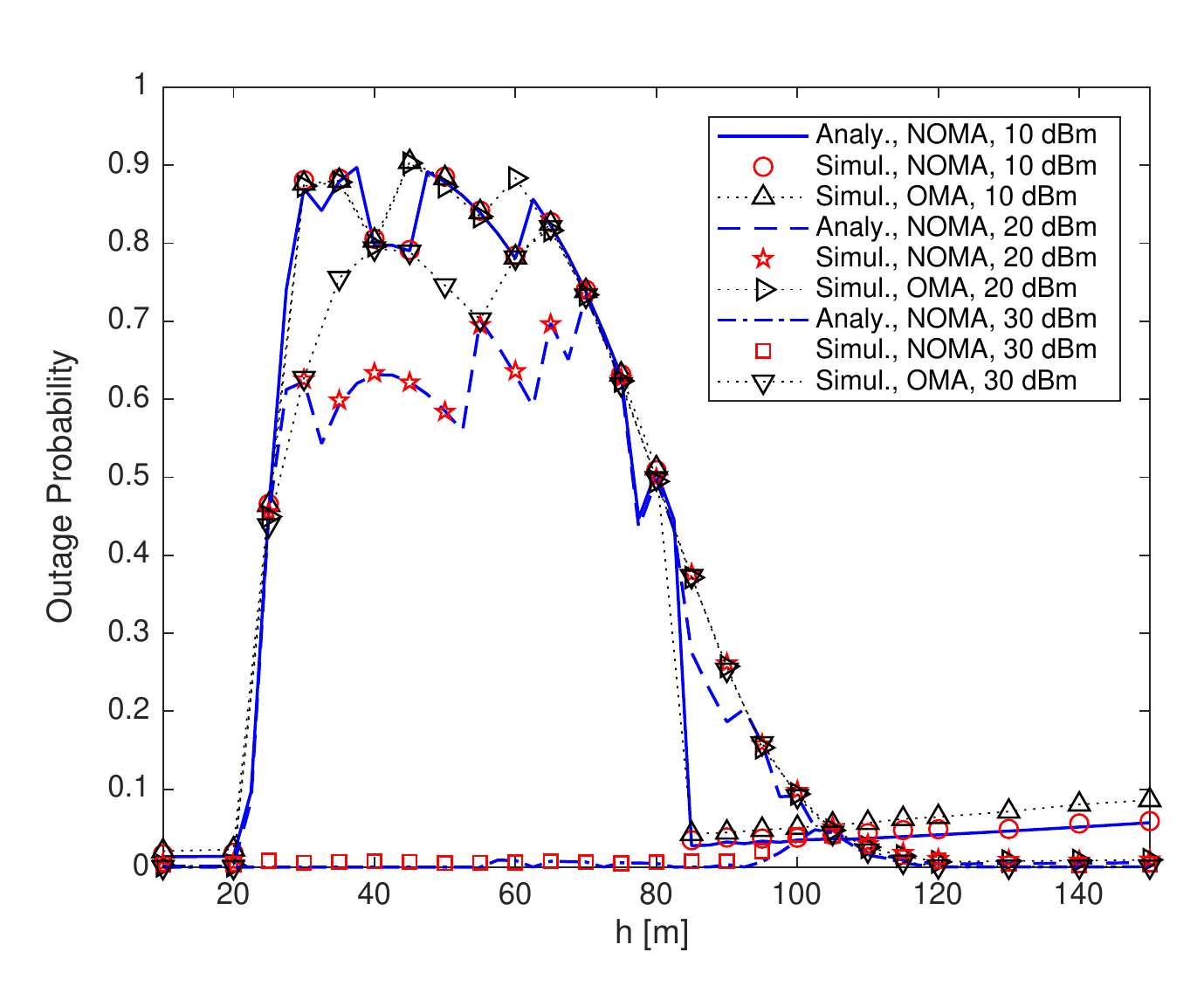}
\label{fig:outage_distance_j20_i30_ith}}
\hspace*{-0.2in}
\subfloat[$j$-th user with $j\,{=}\,20$.]{\includegraphics[width=0.55\textwidth]{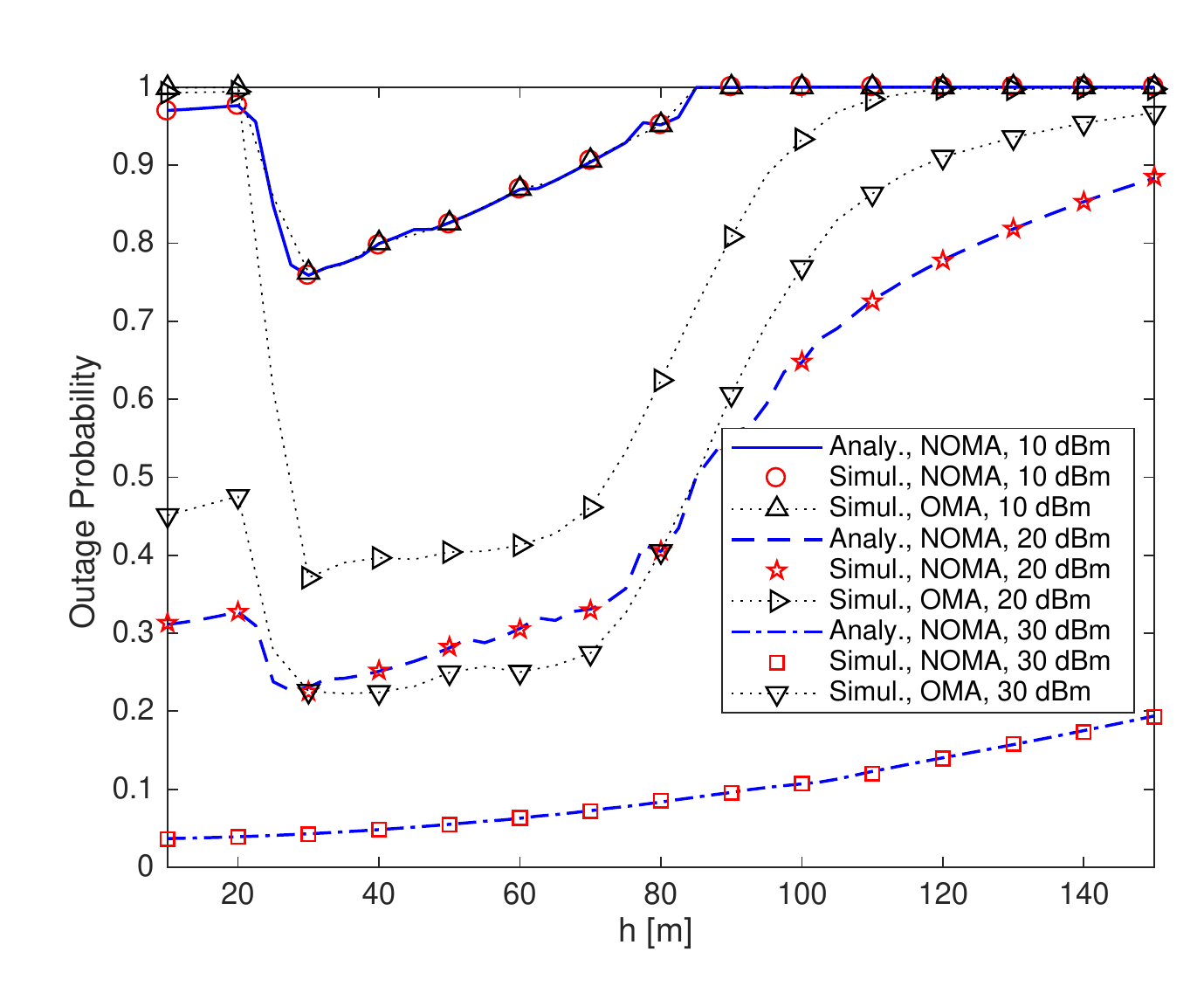}
\label{fig:outage_distance_j20_i30}}
\caption{Outage probability of OMA and NOMA with distance feedback.}
\label{fig:outage_distance_j20_i30}
\end{figure}

\begin{figure}[!t]
\vspace{-2em}
\centering
\includegraphics[width=0.60\textwidth]{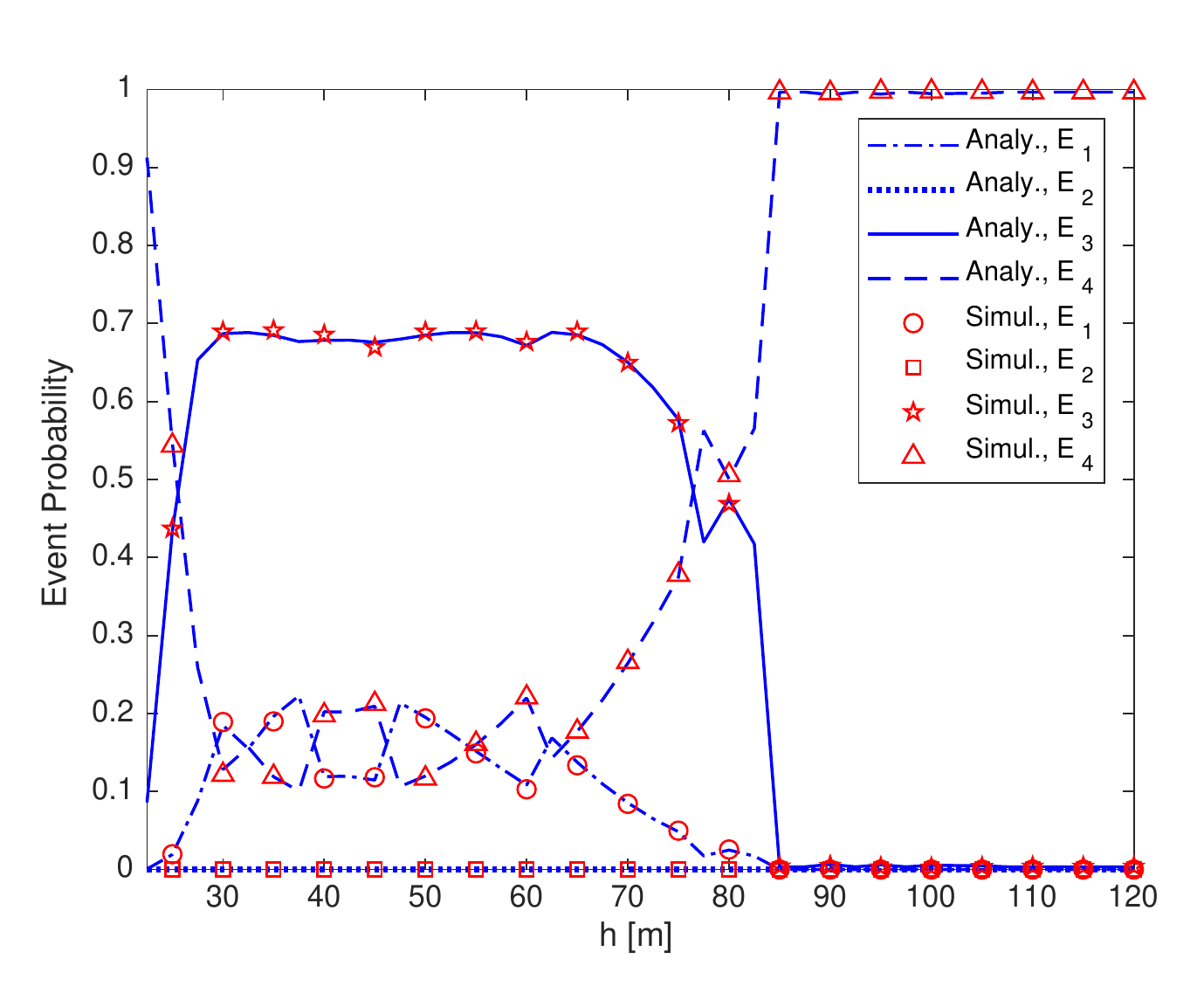}
\caption{Event probability variation for distance feedback scheme over the altitude range corresponding to partially covered user region. At each altitude, radiated region is identified via beam scanning as discussed in Section~\ref{Sec:Beam_Scanning}. Here, $i\,{=}\,30$, $j\,{=}\,20$ and $P_{\rm Tx}=10$~dBm.}
\label{fig:eventprob_j20_i30_varh_optD}
\end{figure}

We observe that sum rates are not monotonic with increasing altitude for the transmit power values of $10$~dBm and $20$~dBm, and, hence, the optimal operation altitude of UAV-BS appears to be $27.5$~m and $25$~m, respectively, for those two transmit power values. When the user region is fully covered ($h\,{\leq}\,20$~m), both $i$-th and $j$-th users are present within the radiated region (${\rm P}\{\textrm{E}_4\} = 1$) and NOMA scheme tries to serve both of them. During this situation and with $P_{\rm Tx}\,{=}\,10$~dBm, the allocated power to $j$-th user is insufficient, and, hence, low target rate of $i$-th user dominates in the sum rates, i.e., at $h\,{=}\,10$~m,  $R^{\textrm{NOMA}}\,{=}\,0.66$~BPCU. On the other hand, when $h\,{\geq}\,20$~m, ${\rm P}\{\textrm{E}_4\}$ starts decreasing rapidly due to the partial coverage of the user region whereas ${\rm P}\{\textrm{E}_3\}$ starts increasing while enhancing the existence probability of only the $j$-th user within the radiated region, as captured in Fig.~\ref{fig:eventprob_j20_i30_varh_optD}. Whenever $j$-th user is scheduled as the only user, single user transmission is realized as described in Section~\ref{sec:noma_events}, and hence the respective outage probability of $j$-th user decreases. The resulting sum rate is then dominated by the target rate of $j$-th user. Due to this reason, we observe a maximum sum rate value at a specific altitude and after that sum rates start decreasing due to the increasing PL and decreasing probability of finding $j$-th user, as shown in Fig.~\ref{fig:eventprob_j20_i30_varh_optD}.

\subsection{Effect of Feedback Type: Distance vs. Full CSI Feedback}\label{sec:numres:feedback_type}
\begin{figure}[!t]
\vspace{-2em}
\centering
\includegraphics[width=0.60\textwidth]{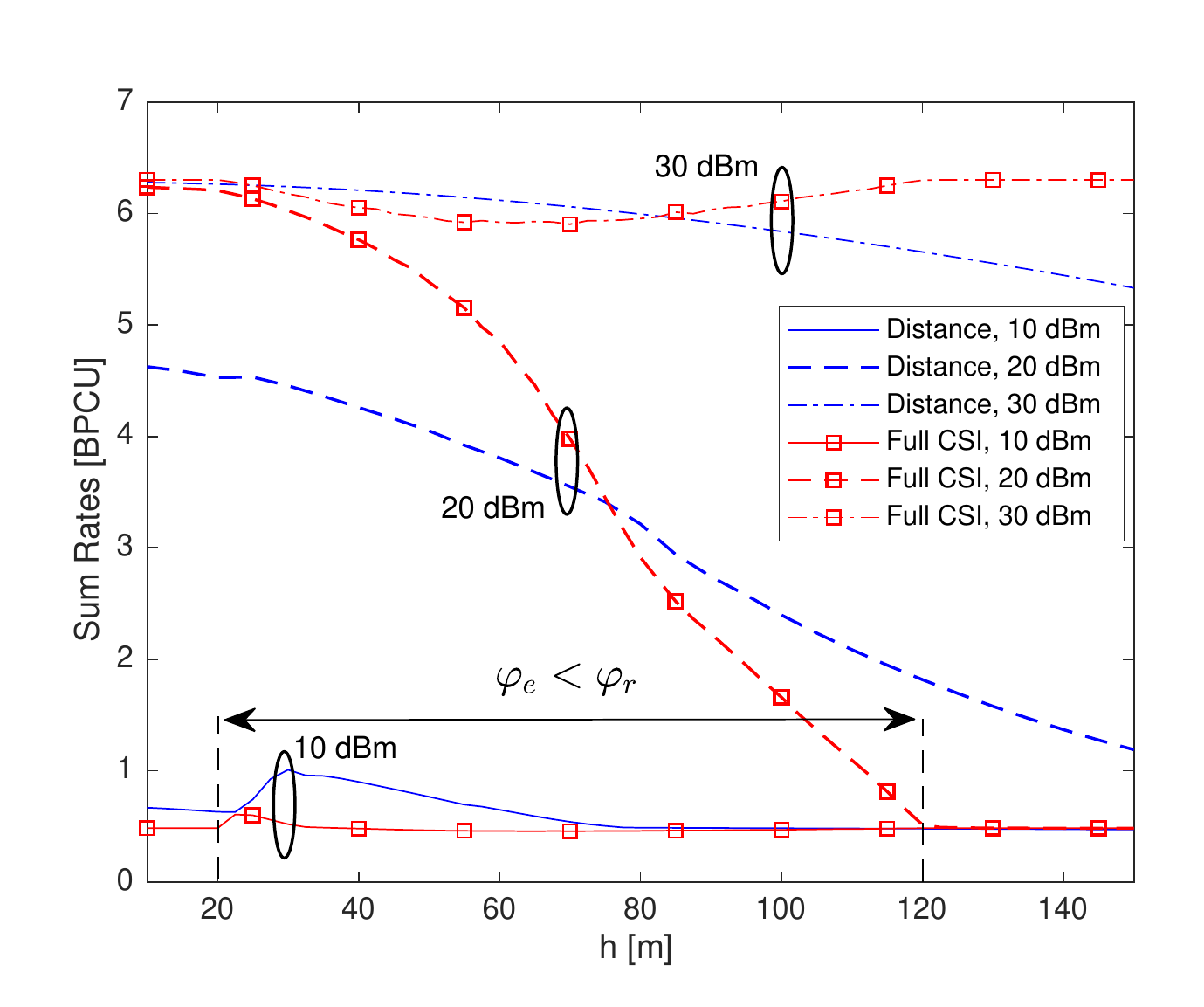}
\caption{Sum rates of NOMA with full CSI and distance feedback where $i\,{=}\,25$, $j\,{=}\,20$. Altitude range for partially covered user region is explicitly shown for which $\varphi_e\,{<}\,\varphi_r$.}
\label{fig:sumrate_distance_vs_fullcsi}
\end{figure}

\begin{figure}[!t]
\vspace{-2em}
\centering
\hspace*{-0.3in}
\subfloat[Full CSI Feedback]{\includegraphics[width=0.53\textwidth]{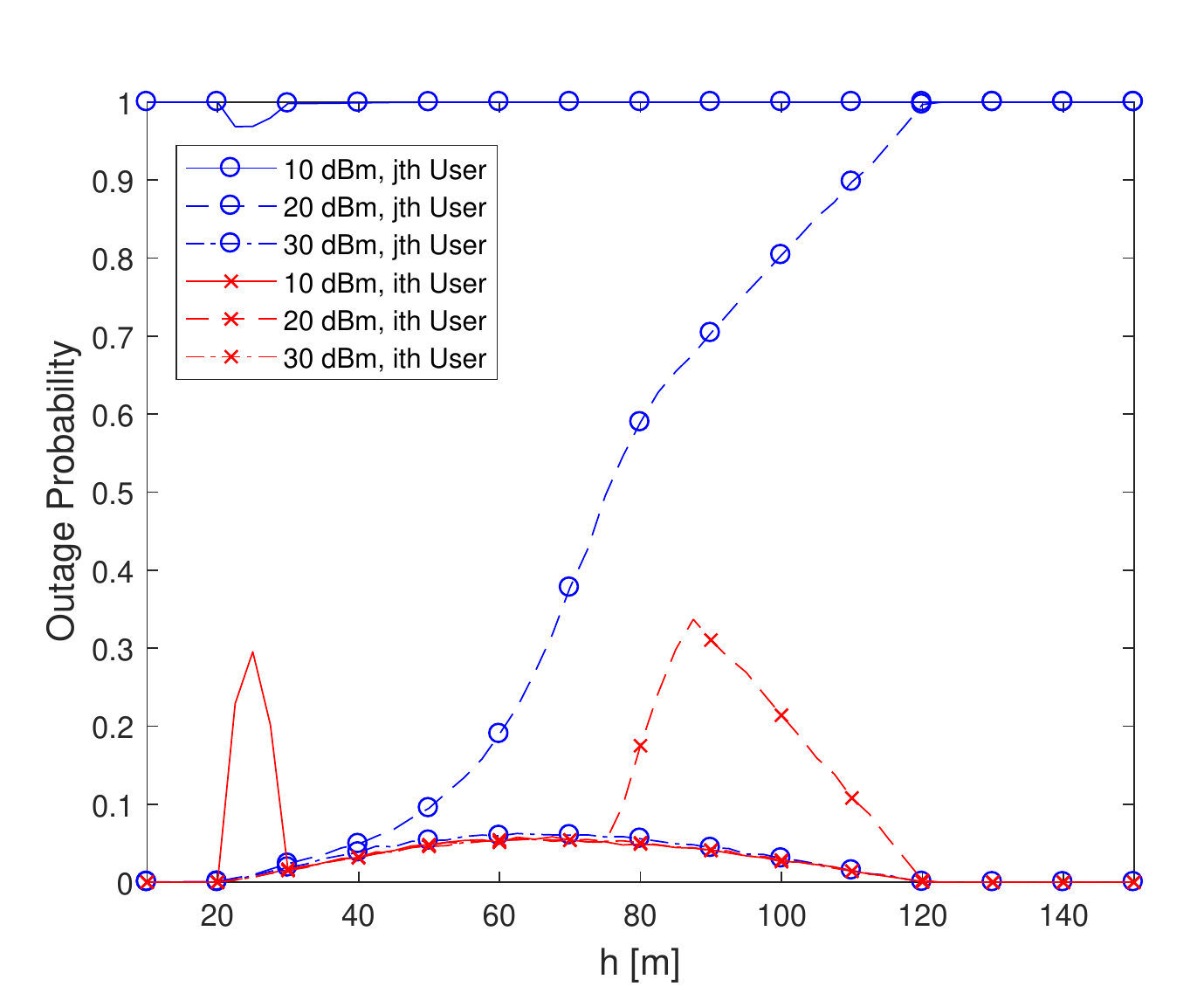}
\label{fig:outage_fullcsi_comparison}}
\hspace*{-0.2in}
\subfloat[Distance Feedback]{\includegraphics[width=0.53\textwidth]{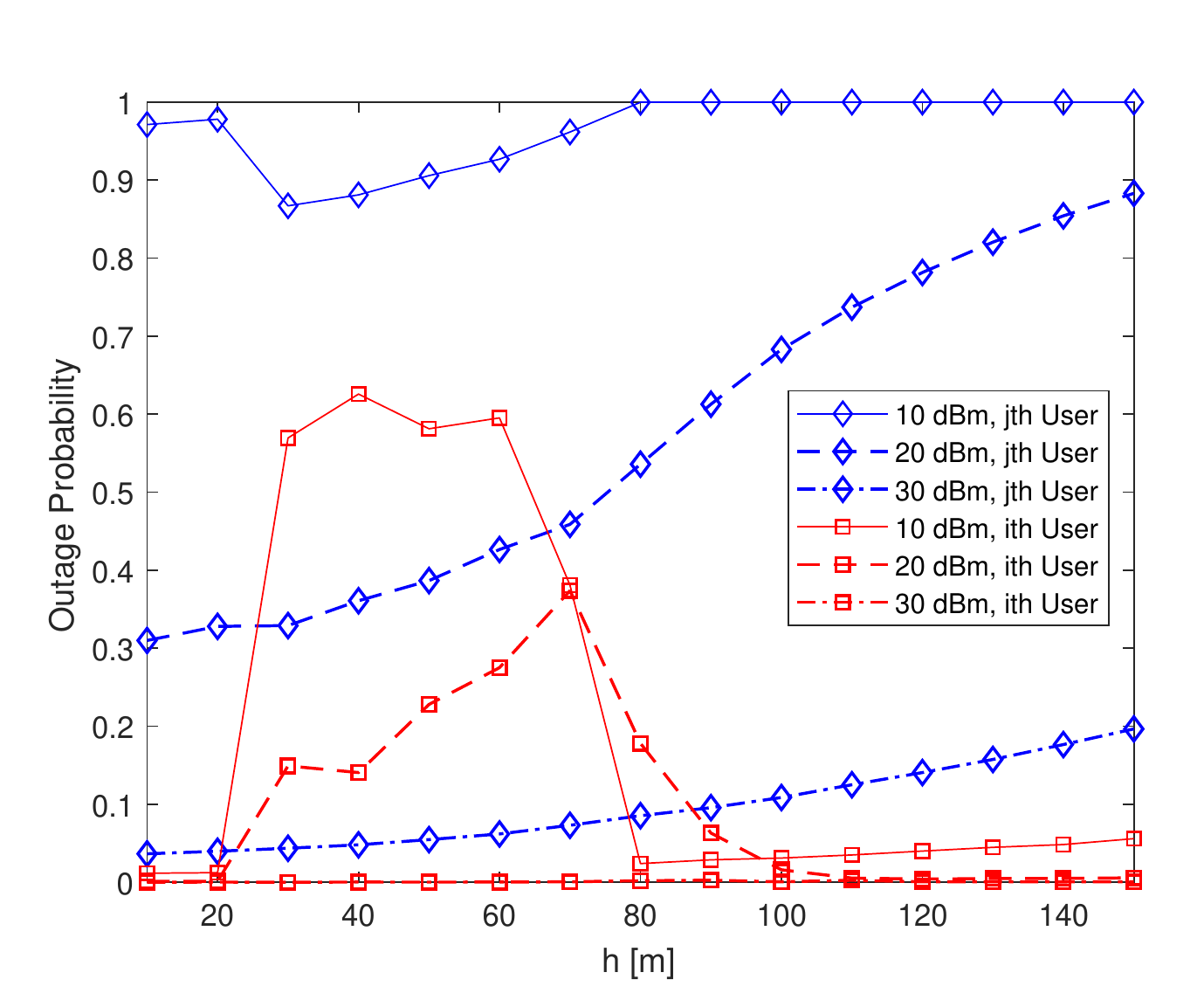}
\label{fig:outage_distance_comparison}}
\caption{Outage of NOMA with full CSI and distance feedback, where $i\,{=}\,25$ and $j\,{=}\,20$ are for distance feedback while $i\,{=}\,K{-}24$ and $j\,{=}\,K{-}19$ with $K\,{=}\,46$ are for full CSI feedback.}
\label{fig:outage_distance_vs_fullcsi}
\end{figure}

In Fig.~\ref{fig:sumrate_distance_vs_fullcsi} and Fig.~\ref{fig:outage_distance_vs_fullcsi}, we depict sum rates and outage performances, respectively, for NOMA with full CSI and distance based feedback schemes. Considering the ordering schemes in Section~\ref{sec:feedback_noma}, full CSI feedback picks $i\,{=}\,K{-}24$ and $j\,{=}\,K{-}19$ while distance feedback assumes $i\,{=}\,25$ and $j\,{=}\,20$, in order to make sure users with the same order of channel quality are selected under two feedback schemes. We observe in Fig.~\ref{fig:sumrate_distance_vs_fullcsi} that sum rates improve along with the increasing transmit power for both feedback schemes, and that the better of these two feedback mechanisms depends on the operation altitude of UAV-BS and the transmit power. In particular, for smaller transmit power, $P_{\rm Tx}\,{=}\,10$~dBm, distance feedback scheme provides better sum rates compared to full CSI up to an altitude of $80$~m. However, there is no performance difference for higher altitudes ($h\,{>}\,80$~m). Note that, for this transmit power, $j$-th user is in complete outage, while $i$-th user has similar outage performance for both feedback schemes at high altitudes, as captured in Fig.~\ref{fig:outage_distance_vs_fullcsi}. At a moderate transmit power of $20$~dBm, full CSI feedback is superior to distance feedback up to $75$~m, and falls short of it after that altitude level. Interestingly, for relatively higher transmit power of $P_{\rm Tx}\,{=}\,30$~dBm, we have an opposite behavior such that full CSI feedback is first inferior to distance feedback for altitudes up to $82.5$~m, and beats it for higher altitudes. Although it will be discussed in more detail later on, we briefly note here that there is an optimal altitude for UAV-BS operation due to the non-monotonic behavior of sum rates along with operational altitudes for certain settings. For example, NOMA with distance feedback and $P_{\rm Tx}\,{=}\,10$~dBm achieves the best sum rate of $1$~BPCU at $h\,{=}\,30$~m.

\begin{figure}[!t]
\vspace{-2em}
\centering
\hspace*{-0.3in}
\subfloat[$i$-th user]{\includegraphics[width=0.53\textwidth]{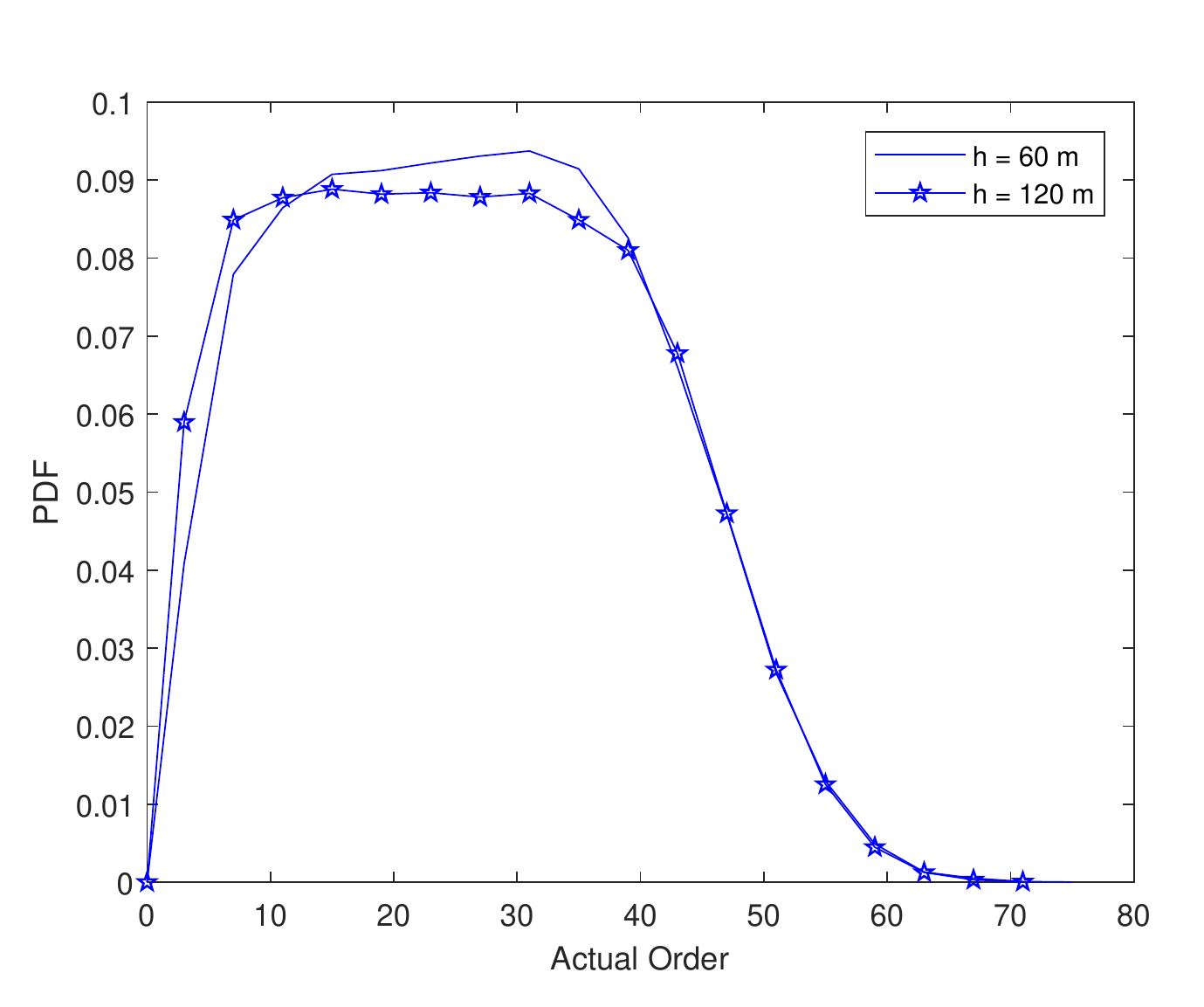}
\label{fig:actual_order_ith}}
\hspace*{-0.2in}
\subfloat[$j$-th user]{\includegraphics[width=0.53\textwidth]{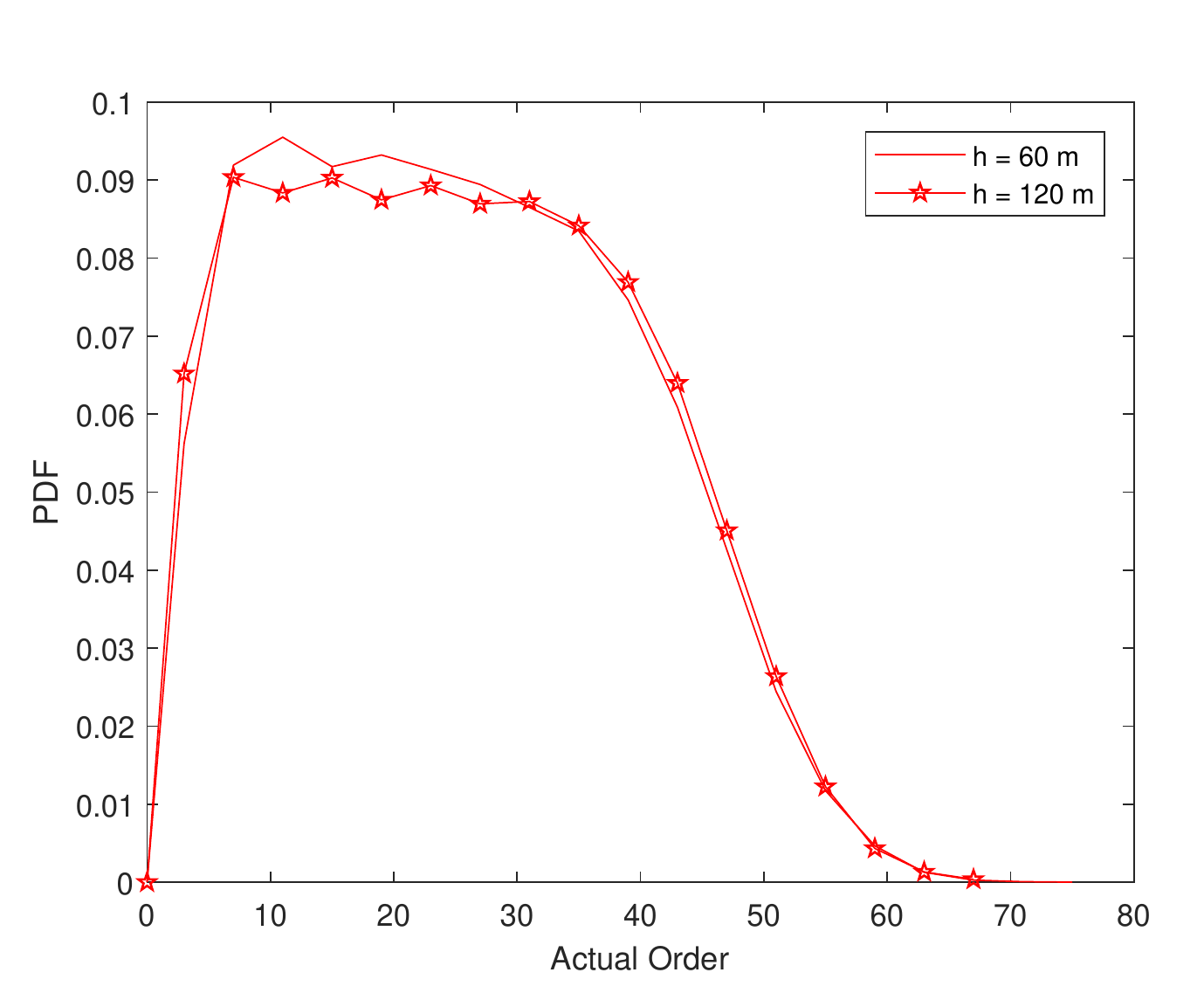}
\label{fig:actual_order_jth}}
\caption{Distribution of actual order for distance feedback with $i\,{=}\,25$, $j\,{=}\,20$, and $h\,{=}\,\{60,120\}$~m.}
\label{fig:actual_order}
\end{figure}

An important point to note here is that, although distance feedback scheme lacks the information of actual channel quality which full CSI feedback exploits completely, its sum rate performance can sometimes be better, as seen in Fig.~\ref{fig:sumrate_distance_vs_fullcsi}, and also mentioned in~\cite{Ding17PoorRandBeamforming, Yapici2018NOMAvlc}. This is because, when  distance based ordering is considered, the order of channel quality with respect to full CSI, or, equivalently, effective channel
gain is not fixed. In particular, distance feedback based ordering picks up users each time with a different \textit{actual order}, which is the desired order with respect to effective channel gain. Since the actual order of NOMA users directly affects outage and sum rates, distance feedback therefore achieves varying performance over trials, which may either be better or worse than that of full CSI based ordering depending on the transmission settings and distribution of actual order. To provide a better insight, we depict the simulation based PDF of actual indices (with respect to the actual order) of NOMA users in Fig.~\ref{fig:actual_order}, whose indices are $i\,{=}\,25$ and $j\,{=}\,20$, for the distance based ordering. We observe that although $i$ and $j$ is fixed for distance based ordering, corresponding actual indices with respect to actual order can take much different values.

\subsection{Effect of Beam Scanning and User Separation}\label{sec:numres:beam_scanning}
\begin{figure}[!t]
\centering
\hspace*{-0.3in}
\subfloat[Sum Rates]{\includegraphics[width=0.55\textwidth]{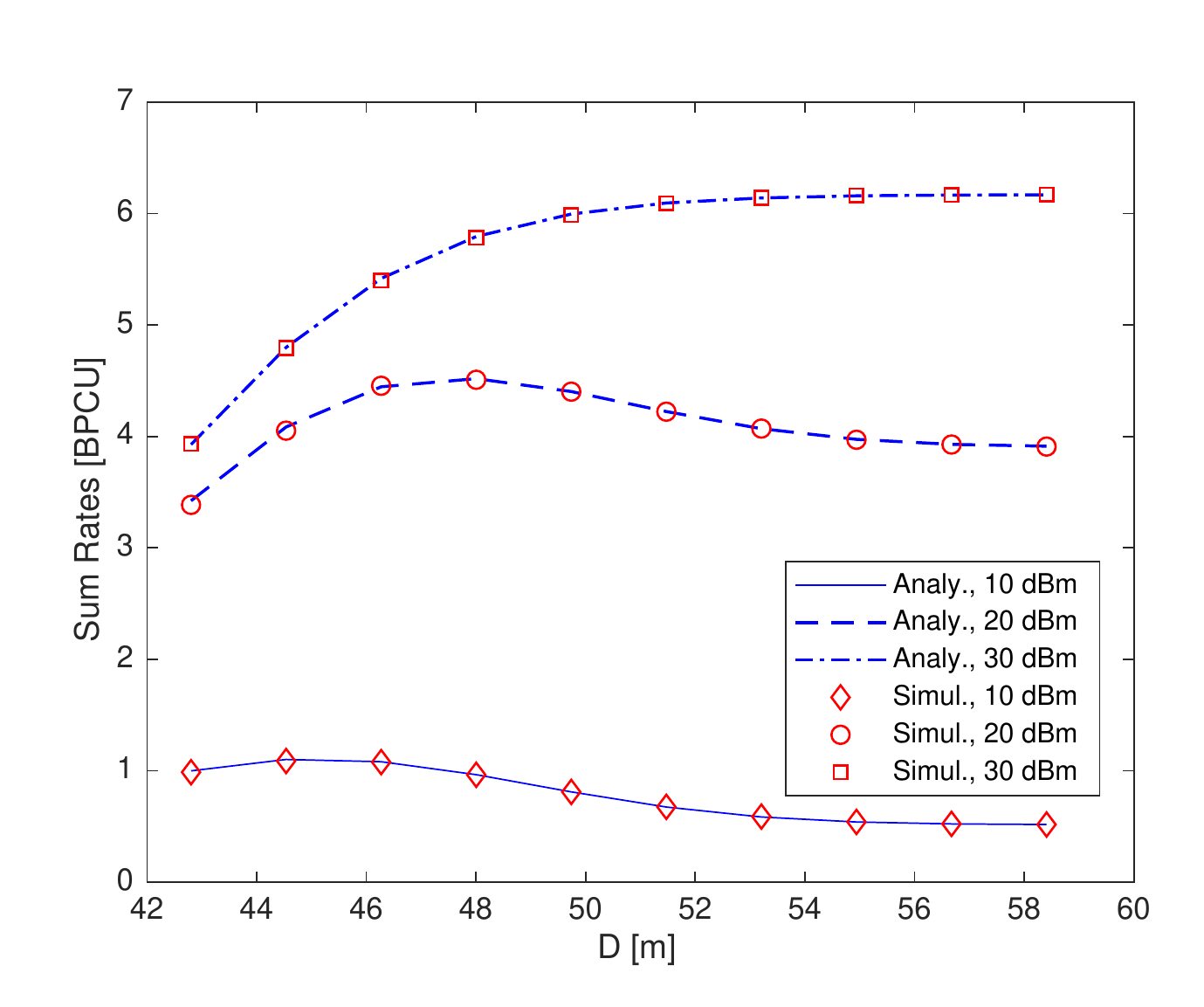}
\label{fig:sumrate_beamscan_j20_i30_h50}}
\hspace*{-0.2in}
\subfloat[Event Probability]{\includegraphics[width=0.55\textwidth]{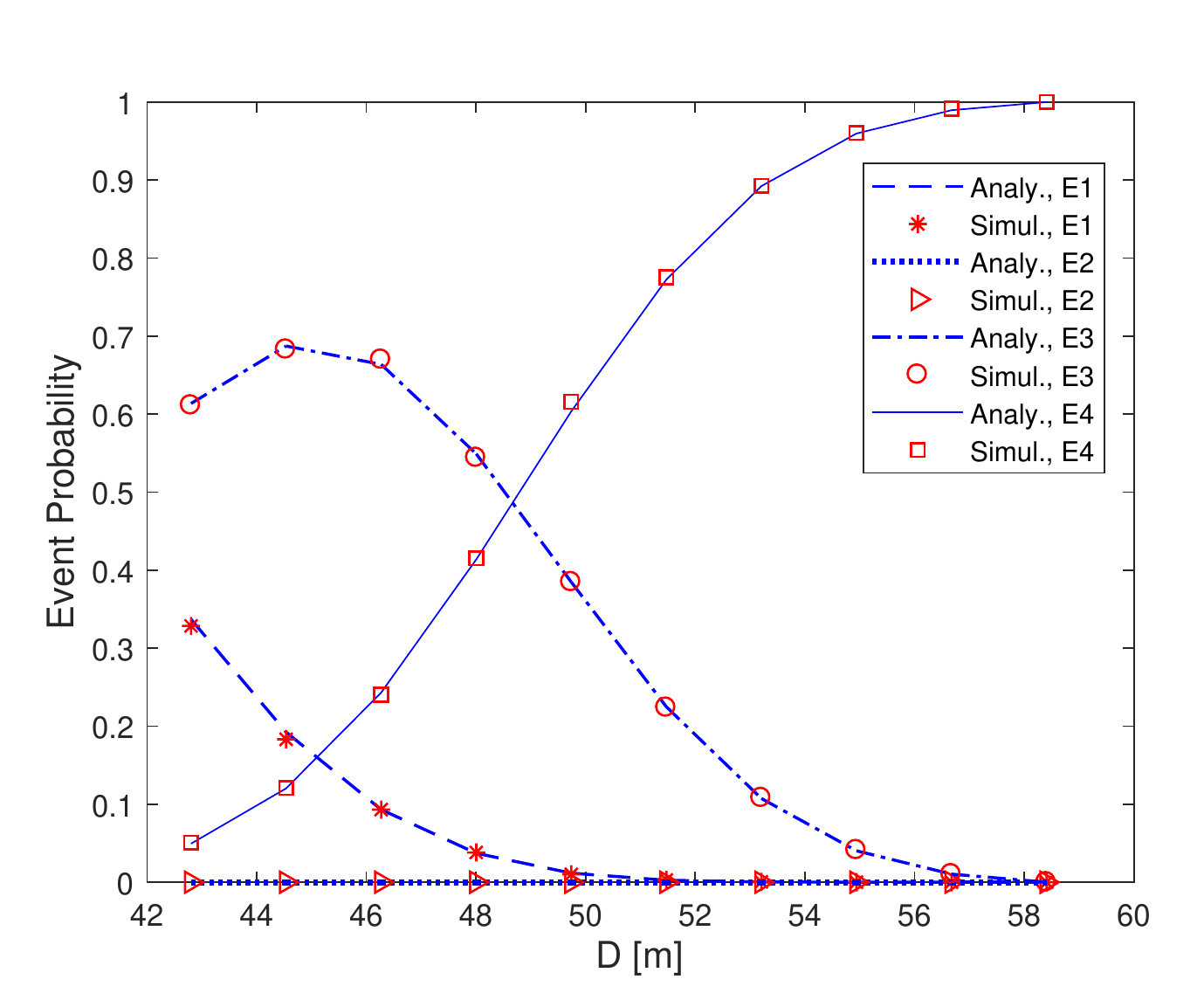}
\label{fig:eventprob_j20_i30_h50_varD}}
\caption{Effect of beam scanning on NOMA sum rates together with event probabilities for $i\,{=}\,30$, $j\,{=}\,20$, and $h\,{=}\,50$~m, where we assume distance feedback scheme.}
\label{fig:beam_scanning}
\end{figure}

In Fig.~\ref{fig:beam_scanning}, we present the impact of beam scanning on the sum rate performance of NOMA with distance feedback along with the respective event probabilities considering $h\,{=}\,50$~m, $i\,{=}\,30$ and $j\,{=}\,20$. As discussed in Section~\ref{Sec:Beam_Scanning}, when $\varphi_r\,{>}\,\varphi_e$, through beam scanning the optimal value of boresight intersection point $D$ is searched to identify the radiated region within the user region. The possible $D$ values vary between $D_1 \,{=}\, 42.8$~m and $D_2 \,{=}\, 58.4$~m to keep the radiated region within the user region boundaries. We observe that the optimal $D$ is $45$~m and $48$~m for the transmit power of $10$~dBm and $20$~dBm, respectively, while any value $D\,{\geq}\,53$~m seems suitable for the optimal operation with $30$~dBm. Note that, conventional NOMA is more likely as $D$ gets larger since finding both users is more probable as can be observed from Fig.~\ref{fig:beam_scanning}\subref{fig:eventprob_j20_i30_h50_varD} increasing event probability $\textrm{E}_4$. In contrast, it is more likely to find only $j$-th user, represented by $\textrm{E}_3$, for relatively smaller $D$ values. As a result, whenever the transmit power is sufficient to serve $j$-th user at least at its target rate, large $D$ values are preferred to benefit from NOMA. On the other hand, smaller $D$ values corresponding to smaller ${\rm P}\{\textrm{E}_4\}$ are better for low transmit power values to leverage transmission only to $j$-th user with full power.

\begin{figure}[!t]
\vspace{-2em}
\centering
\includegraphics[width=0.60\textwidth]{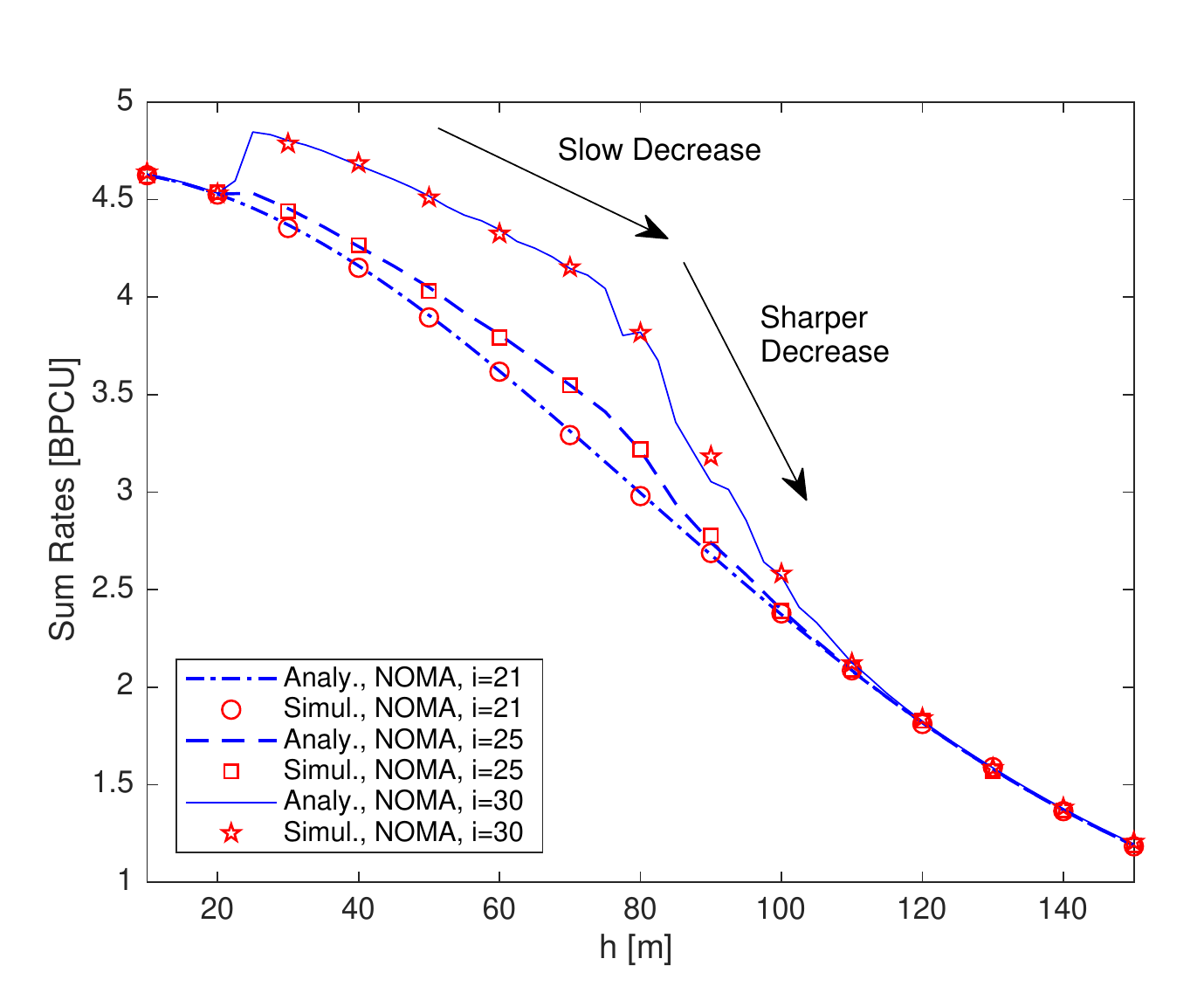}
\caption{Effect of user separation on NOMA sum rates with distance feedback for $i\,{=}\,\{21,25,30\}$, $j\,{=}\,20$, and $P_{\rm Tx}\,{=}\,20$~dBm transmit power.}
\label{fig:sumrate_distance_usersep}
\end{figure}

We consider the effect of user separation after ordering them based on distances, on NOMA sum rates in Fig.~\ref{fig:sumrate_distance_usersep}, assuming $i\,{=}\,\{21,25,30\}$, $j\,{=}\,20$, and transmit power of $20$~dBm. We observe that increasing the user separation $|i{-}j|$ results in larger sum rates over the the altitude range corresponding to partially covered user region. This is because, when the user separation is small, ${\rm P}\{\textrm{E}_4\}$ increases since both users have similar distances and finding both of them within the radiated region is highly probable. This results in serving both of them simultaneously using NOMA. As discussed in Section~\ref{sec:numres:beam_scanning}, with limited transmit power, single user transmission is preferable over NOMA to achieve better sum rates. Hence, better sum rates are observed for larger $i,\, j$ separation here mainly because of the single user transmission due to smaller ${\rm P}\{\textrm{E}_4\}$.

We observe a maximum in sum rates around the altitude of $25$~m for $i\,{=}\,30$, from Fig.~\ref{fig:sumrate_distance_usersep}. As discussed previously and can be seen from Fig.~\ref{fig:eventprob_j20_i30_varh_optD}  for ${\rm P}\{\textrm{E}_3\}$, this is because, there is a higher chance of scheduling only the $j$-th user up to the altitude of $80$~m. Due to the increasing PL, sum rates start to drop with a slower rate till $80$~m. After $80$~m sum rates drop with a higher rate due to the increasing of ${\rm P}\{\textrm{E}_4\}$ as captured in Fig.~\ref{fig:eventprob_j20_i30_varh_optD} in addition to PL.



\subsection{NOMA Performance with mmWave PL Model}\label{sec:numres:noma_mmW}
\begin{figure}[!t]
\vspace{-2em}
\centering
\includegraphics[width=0.60\textwidth]{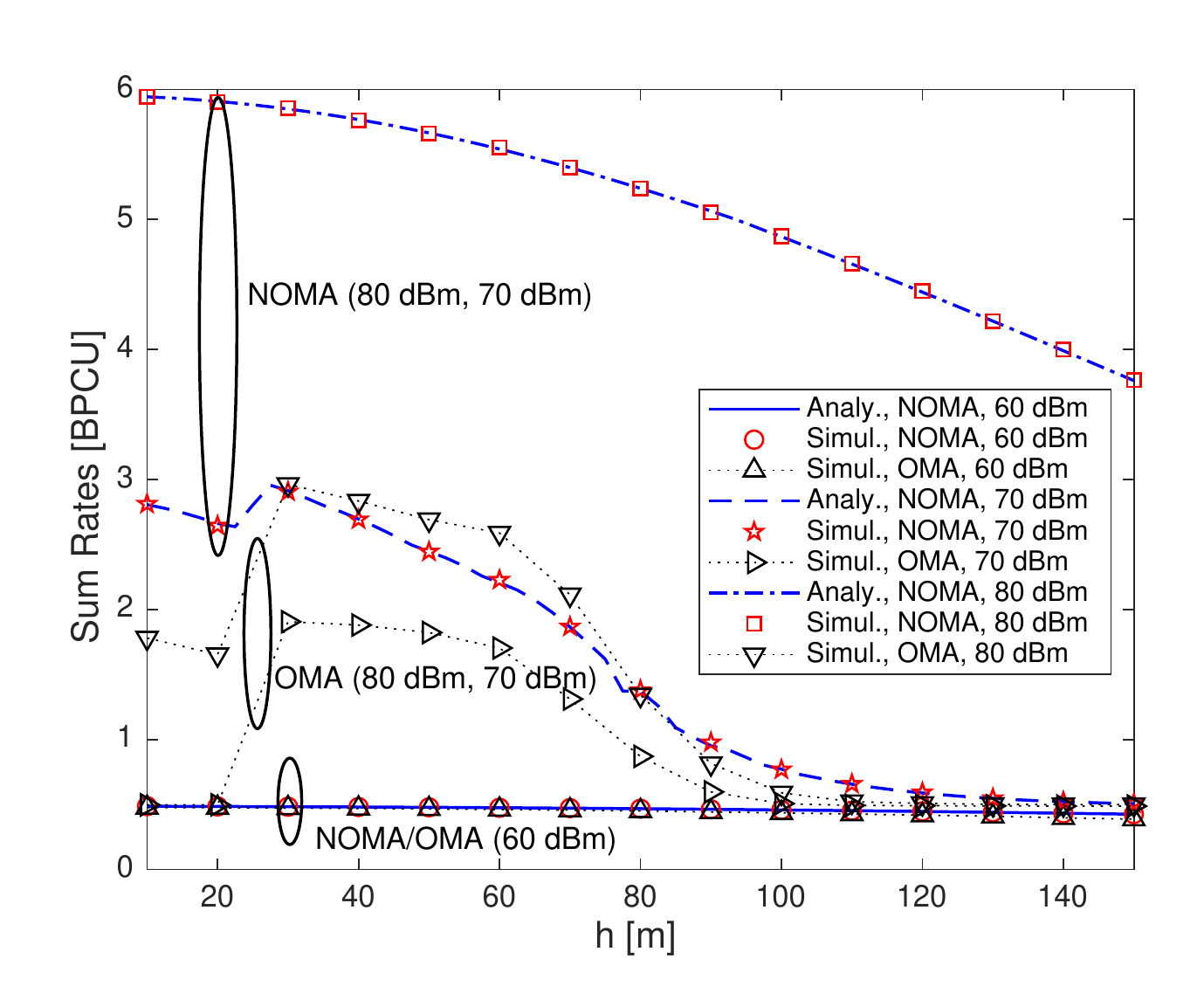}
\caption{Sum rates of OMA and NOMA with distance feedback for CI mmWave channel where $i\,{=}\,25$, $j\,{=}\,20$.}
\label{fig:sumrate_distance_j20_i25_mmW}
\end{figure}

\begin{figure}[!t]
\vspace{-2em}
\centering
\hspace*{-0.3in}
\subfloat[$i$-th user with $i\,{=}\,25$]{\includegraphics[width=0.55\textwidth]{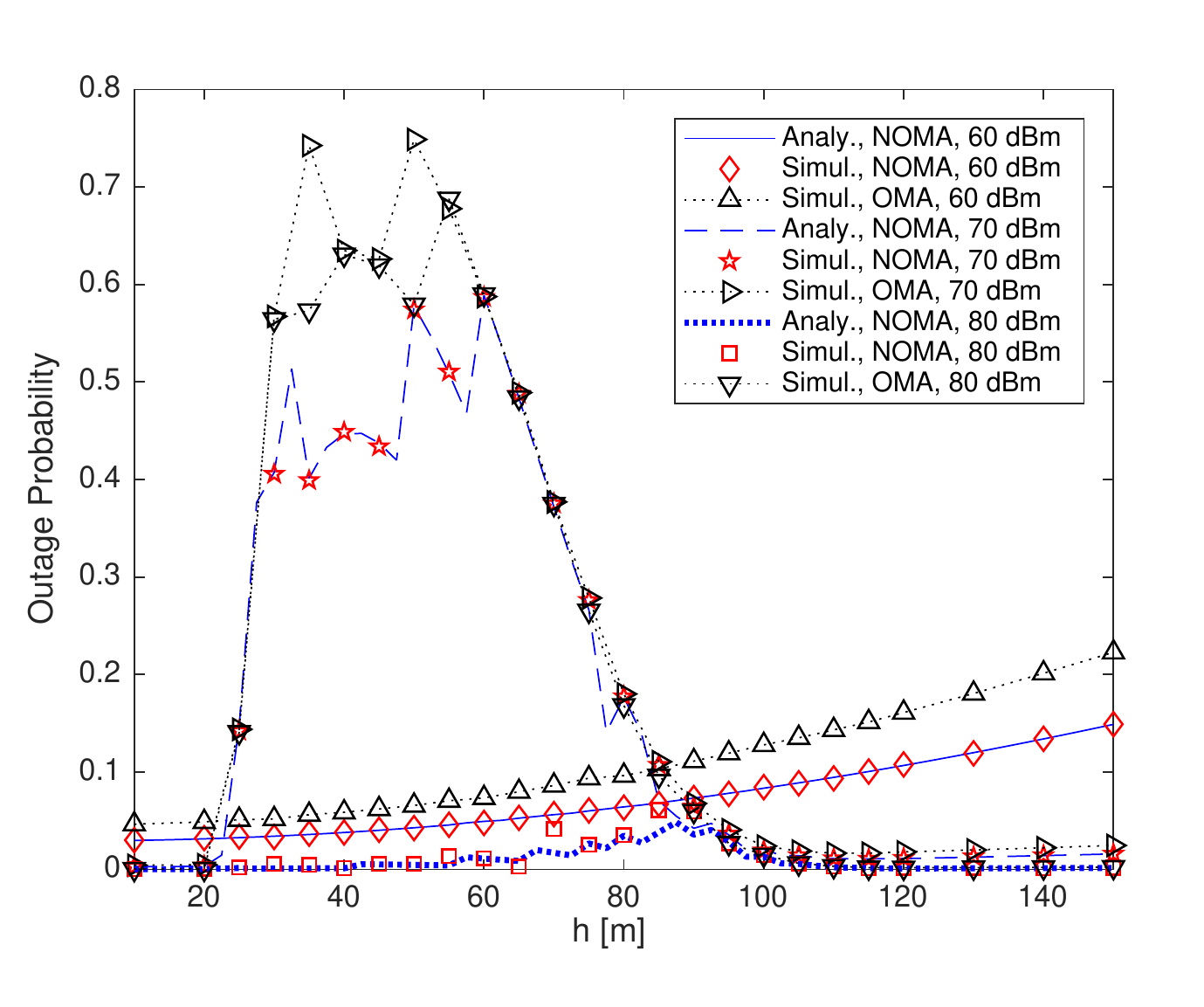}
\label{fig:outage_distance_j20_i25_ith_mmW}}
\hspace*{-0.2in}
\subfloat[$j$-th user with $j\,{=}\,20$]{\includegraphics[width=0.55\textwidth]{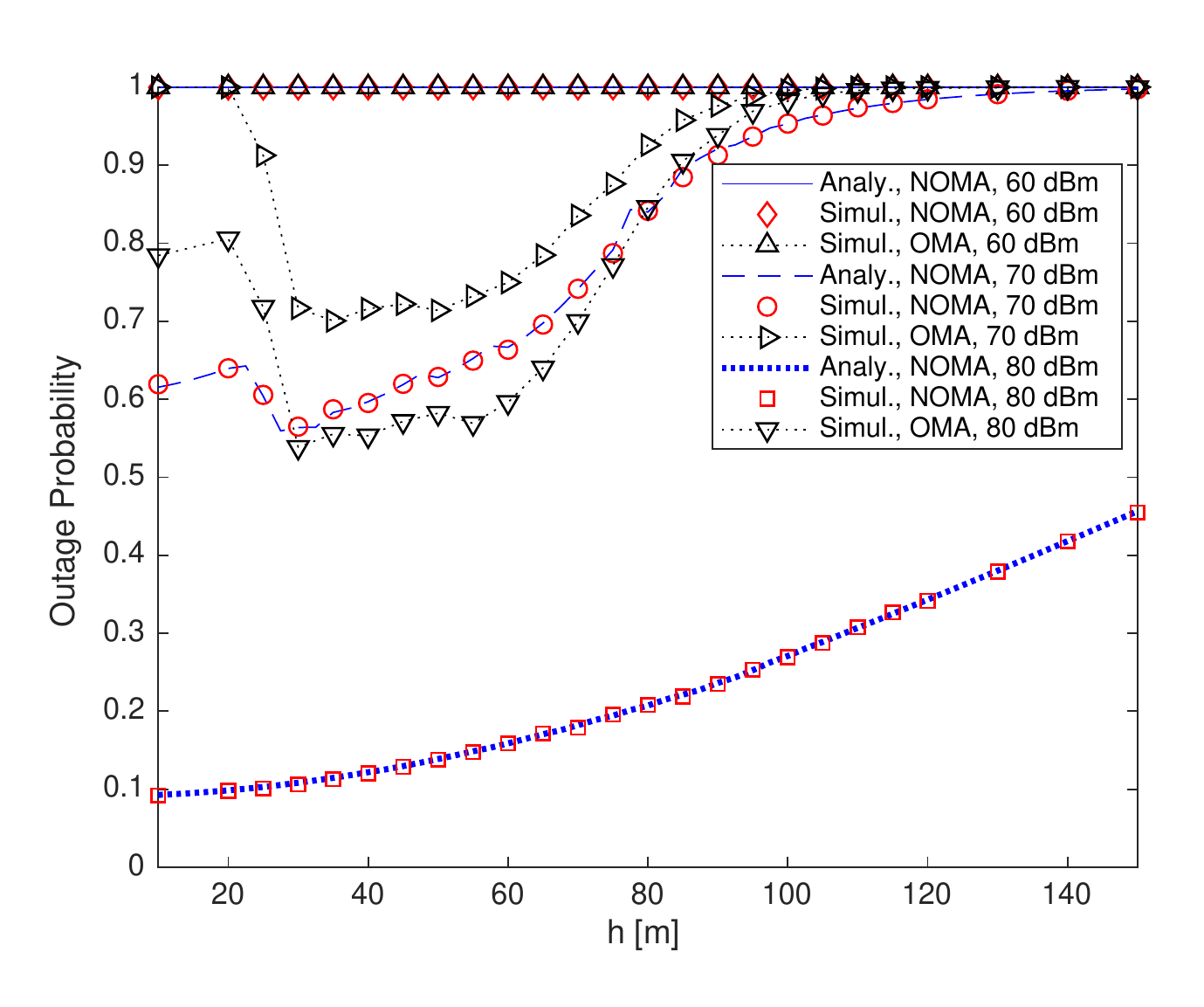}
\label{fig:outage_distance_j20_i25_jth_mmW}}
\caption{Outage probability of OMA and NOMA with distance feedback for CI mmWave channel.}
\label{fig:outage_distance_j20_i25_mmW}
\end{figure}

In this section, we investigate achievable sum rates and outage performance for OMA and NOMA with distance feedback scheme considering the CI mmWave path-loss model defined previously, and depict the respective results in Fig.~\ref{fig:sumrate_distance_j20_i25_mmW} and Fig.~\ref{fig:outage_distance_j20_i25_mmW}, for $i\,{=}\,25$ and $j\,{=}\,20$. From Fig.~\ref{fig:sumrate_distance_j20_i25_mmW} we observe that analytical sum rate results for NOMA perfectly match with the simulation based results, as before, and that NOMA results are better than OMA except at $P_{\rm Tx}\,{=}\,60$~dBm. Due to the severe PL in mmWave frequency bands, the required transmit power level is observed to be relatively larger. It is worth remarking that, we consider $M\,{=}\,100$, since it is reasonable to expect more antenna elements in the array for mmWave frequencies. This will provide a larger beamforming gain to partially compensate the severe PL. Similar to distance dependent path-loss model, with CI PL model also sum rates exhibit a maximum at the altitude of around $27.5$~m for $P_{\rm Tx}\,{=}\,70$~dBm which is accompanied by the decrease in $j$-th user outage probabilities.

\subsection{Asymptotic Behavior of Sum rates and Outage Probabilities}
\begin{figure}[!t]
\vspace{-2em}
{
\centering
\hspace*{-0.3in}
\subfloat[Sum rates.]{\includegraphics[width=0.55\textwidth]{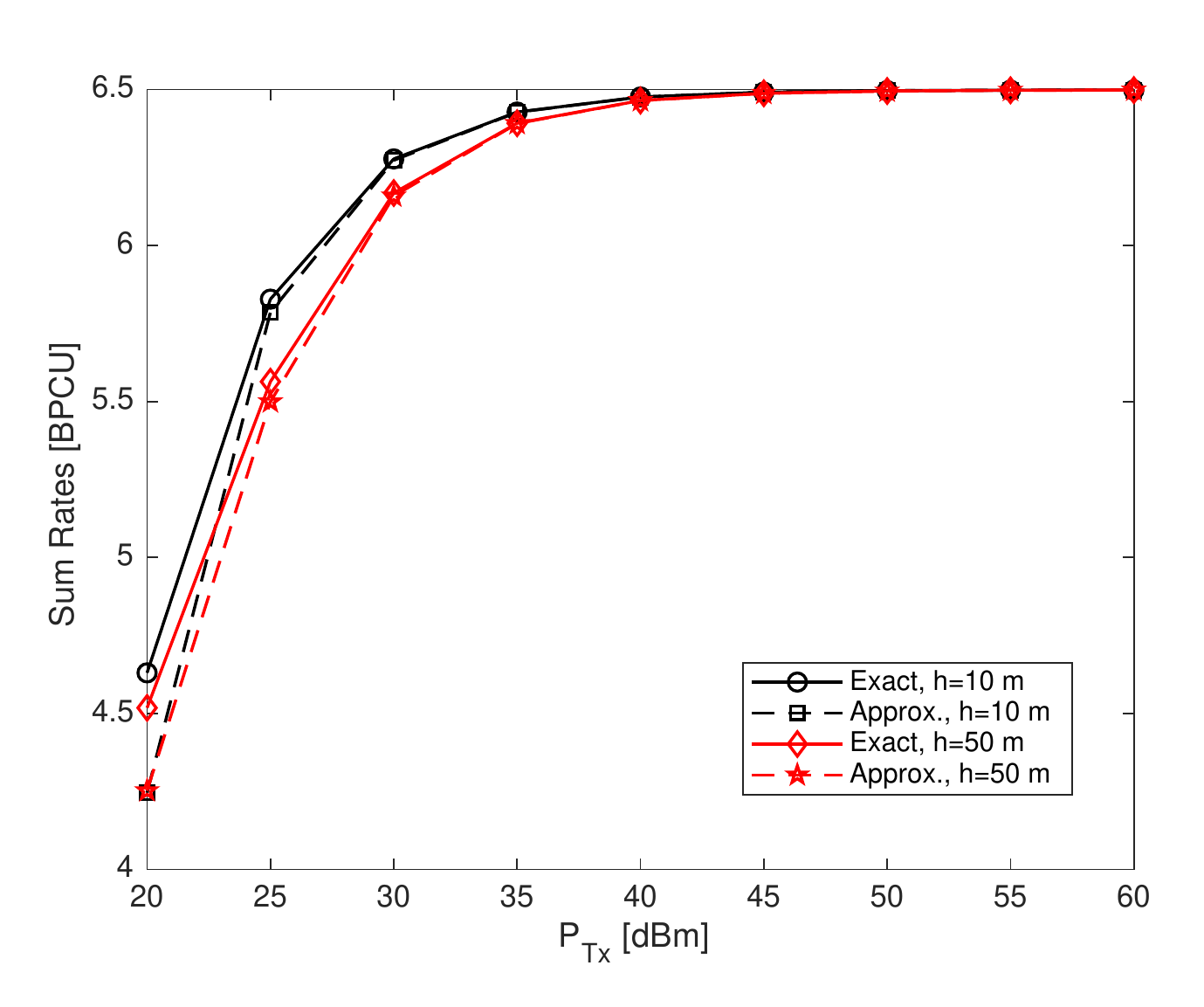}
\label{fig:Asymptotic_sum_rates}}
\hspace*{-0.2in}
\subfloat[$j$-th user outage with $j\,{=}\,20$.]{\includegraphics[width=0.55\textwidth]{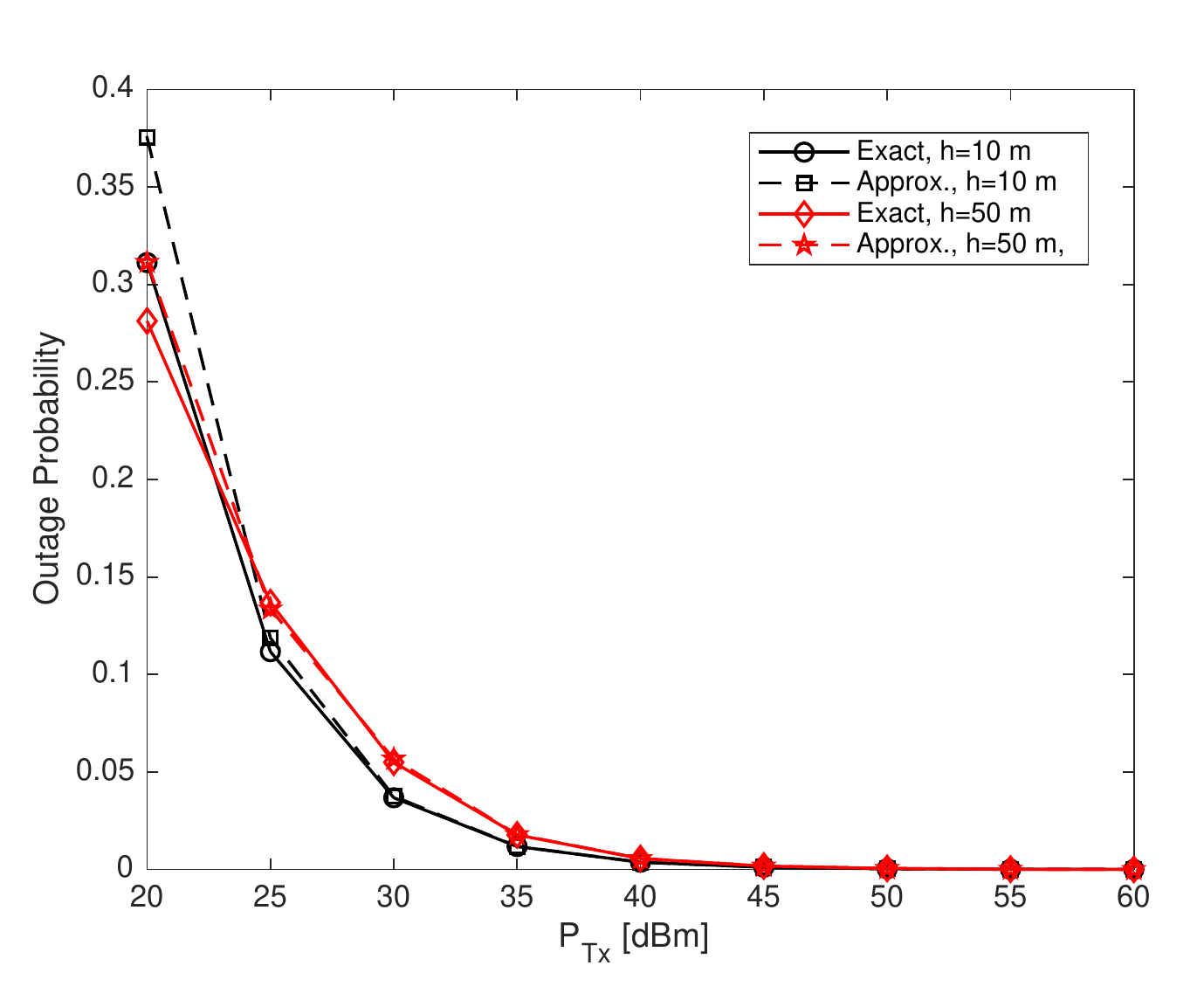}
\label{fig:Asymptotic_outage_j}}
\caption{Asymptotic behavior of sum rates and outage probabilities where $i\,{=}\,30$, $j\,{=}\,20$. Here ``Exact'' results are generated using \eqref{eq:cond_outage_j_sk1_3} and \eqref{eq:cond_outage_kn_sk2_3}, where as ``Approx.'' results are generated considering \eqref{eq:Asy_outage_Sk1} and \eqref{eq:Asy_outage_Sk2} }
\label{fig:Asymptotic_analysis_j20_i30} }
\end{figure}

In Fig.~\ref{fig:Asymptotic_analysis_j20_i30} sum rates and $j$-th user outage probabilities calculated considering exact outage probability expressions of \eqref{eq:cond_outage_j_sk1_3} and \eqref{eq:cond_outage_kn_sk2_3} and approximated asymptotic outage probability expressions of \eqref{eq:Asy_outage_Sk1} and \eqref{eq:Asy_outage_Sk2} are presented for $h=10,\, 50$~m. We observe that exact and asymptotic results are  matching adequately at high SNR (represented by high ${P}_{\rm Tx}$), while there is a gap between them at low SNR, as expected.

\section{Concluding Remarks}\label{sec:conclusion}
In this paper, we introduce NOMA transmission to a UAV-aided communication network which is deployed to provide coverage over a densely packed user region such as a stadium or a concert area. We show that the user region may not be covered completely at particular UAV-BS operational altitudes of practical relevance. During such situations a beam scanning approach is proposed to identify the optimal area to be radiated within the user region. We accordingly propose a hybrid transmission strategy serving all or some of the desired users at a time, which leverages the presence of desired NOMA users in the radiated region.

As a practical feedback mechanism, we consider the availability of user distances as the feedback and subsequently use that for user ordering during NOMA formulation. Interestingly, distance feedback appears to be an efficient alternative for full CSI feedback especially for rapidly fluctuating channels. Our analysis shows that NOMA outperforms OMA even when distance feedback rather than full CSI feedback is utilized. Further, the achievable sum rates are not monotonic with increasing altitudes specifically at smaller transmit power values and there is an optimal hovering altitude for UAV-BS to maximize sum rates. We also shed light on the dependency of achievable sum rates on user selection for NOMA transmission. In particular, we identify that if two users with larger separation is selected for NOMA transmission after distance based ordering, better sum rates can be observed especially over the altitude range corresponding to partially covered user region. Further, by considering CI mmWave path loss model we evaluate achievable sum rates which again proves the dominance of NOMA over its orthogonal counterpart. Finally, through asymptotic analysis we show that achievable diversity gain for NOMA with distance feedback is independent of user index.
\appendices
\begin{figure}
\centering
\includegraphics[width=0.4\textwidth]{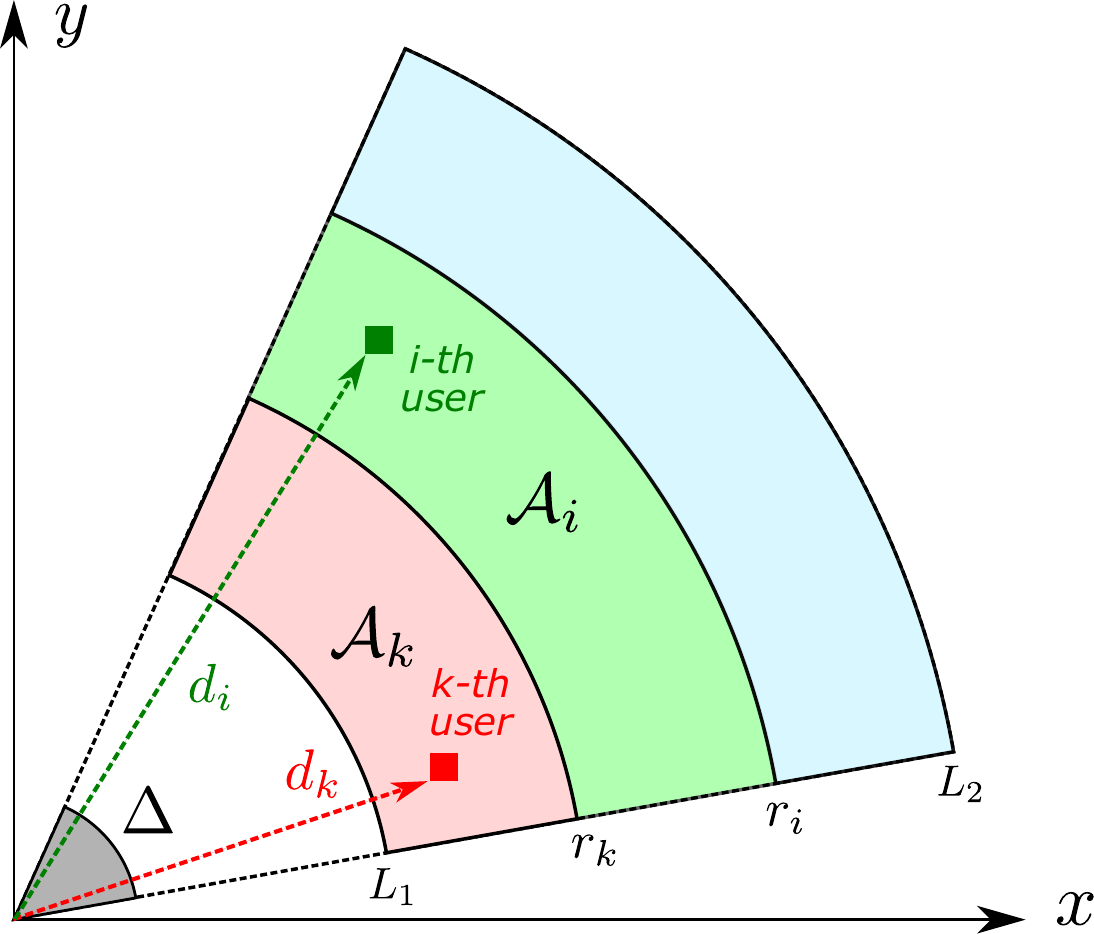} \vspace{1em}
\caption{Sketch of user region $\mathcal{A} \,{=}\, \Delta(L_2^2 \,{-}\, L_1^2)$ with $\mathcal{A}_k \,{=}\, \Delta(r_k^2 \,{-}\, L_1^2)$ and $\mathcal{A}_i \,{=}\, \Delta(r_i^2 \,{-}\, r_k^2)$.}
\label{fig:User_Region}
\end{figure}
\section{The Marginal PDF of User Distances for $\mathcal{S}_{K_1}$} \label{app:PDF_j_less_K_less_i}
Let us first consider the cumulative distribution function (CDF) of the $k$-th user distance $d_k$ assuming that $K$ is chosen from $\mathcal{S}_{K_1}{:} \{ K | j \,{\leq}\, K \,{<}\, i\}$, which is given as
\begin{align}\label{eq:CDF_j_less_K_less_i}
F_{d_k|\mathcal{S}_{K_1}} (r_k) = {\rm P}\{d_k < r_k | \, j \,{\leq}\, K \,{<}\, i \} = \frac{{\rm P}\{ d_k < r_k, \ j \,{\leq}\, K \,{<}\, i \} }{{\rm P}\{ j \,{\leq}\, K \,{<}\, i \} }.
\end{align}
Note that while the denominator of \eqref{eq:CDF_j_less_K_less_i} is readily available from the definition of HPPP, we will relate the ordered user distances to the number of users with a help of Fig.~\ref{fig:User_Region}, as discussed in \cite{Haenggi05Stochastic_Geo}, in order to calculate the probability in the numerator. To this end, the first condition $d_k\,{<}\,r_k$ in the numerator of \eqref{eq:CDF_j_less_K_less_i} is interpreted as the necessity of the area $\mathcal{A}_k$ having at least $k$ users. In addition, given that the number of users in $\mathcal{A}_k$ is $l\,{\geq}\,k$, the second condition $j \,{\leq}\, K \,{<}\, i$ requires that the remaining area $\mathcal{A}\,{-}\,\mathcal{A}_k$ has at most $(i\,{-}\,l\,{-}\,1)$ users, so that the user region has less than $i$ users. As a result, the desired probability is calculated as
\begin{align}
\hspace{-0.14in}{\rm P}\{ d_k \,{<}\, r_k, j \,{\leq}\, K \,{<}\, i \} &\,{=} \sum \limits_{l=k}^{i-1} {\rm P}  \{\mathcal{A}_k \, \textrm{has $l$ users}, \, \mathcal{A}{-}\mathcal{A}_k \, \textrm{has at most $(i\,{-}\,l\,{-}\,1)$ users}\} \nonumber \\
&= \sum \limits_{l=k}^{i{-}1} \frac{e^{{-}\Delta(r_k^2{-}L_1^2)\lambda}\left[ \Delta(r_k^2{-}L_1^2)\lambda \right]^l}{l!}\left\lbrace \sum \limits_{l'=0}^{i{-}l{-}1} \frac{e^{{-}\Delta(L_2^2{-}r_k^2)\lambda}\left[ \Delta(L_2^2{-}r_k^2)\lambda \right]^{l'}}{l'!}\right\rbrace. \label{eq:marginal_cdf}
\end{align}
Using \eqref{eq:marginal_cdf} and $\mathcal{C} \,{=}\, {\rm P}\{ j \,{\leq}\, K \,{<}\, i \} \,{=}\, \sum\limits_{l=j}^{i{-}1} \frac{e^{{-}\Delta(L_2^2{-}L_1^2)\lambda} \left[\Delta(L_2^2{-}L_1^2)\lambda \right]^l}{l!}$, the marginal CDF in \eqref{eq:CDF_j_less_K_less_i} can be readily obtained. Finally, the marginal PDF can be obtained by taking derivative as follows
\begin{align}
f_{d_k|\mathcal{S}_{K_1}} (r_k)&=\frac{\partial}{\partial r_k }F_{d_k|\mathcal{S}_{K_1}} (r_k) \nonumber \\
&=\frac{e^{-\Delta(L_2^2 - L_1^2)\lambda}}{\mathcal{C}}\frac{\partial }{\partial r_k } \left\lbrace  \sum \limits_{l=k}^{i-1} \sum \limits_{l'=0}^{i-l-1} \frac{\left[ \Delta(r_k^2 - L_1^2)\lambda \right]^l}{l!} \frac{\left[ \Delta(L_2^2 - r^2)\lambda \right]^{l'}}{l'!}\right\rbrace \nonumber \\
&=\frac{(2\Delta \lambda r_k)e^{-\Delta(L_2^2 - L_1^2)\lambda} }{\mathcal{C}}\frac{\left[\Delta(r_k^2 - L_1^2)\lambda \right]^{(k-1)}}{(k-1)!} \left( \sum \limits_{l=0}^{i-k-1}\frac{\left[\Delta(L_2^2 - r_k^2)\lambda \right]^{l}}{l!} \right). \qquad \;\; \IEEEQEDhere
\end{align}

\section{The Joint PDF of User Distances for $\mathcal{S}_{K_2}$} \label{app:JPDF_K_greater_i}
Similar to the marginal PDF derivation in Appendix~\ref{app:PDF_j_less_K_less_i}, we first consider the joint CDF of the user distances $d_k$ and $d_i$ with $d_k \leq d_i$ assuming that $K{\in}\mathcal{S}_{K_2}$ with $\mathcal{S}_{K_2}{:} \{ K | K \,{\geq}\, i\}$, which is given as
\begin{align}\label{eq:JCDF}
F_{d_k,d_i|\mathcal{S}_{K_2}} (r_k, r_i) \,{=}\, {\rm P}\{d_k < r_k, d_i < r_i, d_k \leq d_i| K \geq i \}  \,{=}\, \frac{{\rm P}\{ d_k < r_k, d_i < r_i, d_k \leq d_i, K \geq i \} }{{\rm P}\{ K \geq i \} }.
\end{align}

Considering the probability expression in the numerator of \eqref{eq:JCDF} and Fig.~\ref{fig:User_Region}, $\mathcal{A}_k$ should have at least $k$ users to satisfy $d_k \,{<}\, r_k$. In addition, the condition $d_i \,{<}\, r_i$ requires the presence of at least $i$ users in $\mathcal{A}_k+\mathcal{A}_i$, which also meets the condition $ K \,{\geq}\, i $. Given that the number of users in $\mathcal{A}_k$ is $l\,{\geq}\,k$, $\mathcal{A}_i$ has at least $(i-l)$ users. Note that, the maximum number of users in $\mathcal{A}_k$ should be $(i-1)$ to satisfy $d_k \,{\leq}\, d_i$. The desired probability is accordingly given as
\begin{align}
&{\rm P}\{ d_k \,{<}\, r_k, d_i \,{<}\, r_i, d_k \,{\leq}\, d_i, K \,{\leq}\, i \} = \sum \limits_{l=k}^{i-1} {\rm P}  \{\mathcal{A}_k \, \textrm{has at least $l$ users}, \, \mathcal{A}_i \, \textrm{has at least $(i-l)$ users}\} \nonumber \\
& \qquad \qquad = \sum \limits_{l=k}^{i-1} \frac{e^{-\Delta(r^2 - L_1^2)\lambda}\left[ \Delta(r^2 - L_1^2)\lambda \right]^l}{l!}\left\lbrace 1 - \sum \limits_{l'=0}^{i-l-1} \frac{e^{-\Delta(r_i^2 - r^2)\lambda}\left[ \Delta(r_i^2 - r^2)\lambda \right]^{l'}}{l'!}\right\rbrace \label{eq:JCDF_proof_step_3}.
\end{align}
Employing \eqref{eq:JCDF_proof_step_3} and $\mathcal{C} \,{=}\, {\rm P}\{ K \geq i \} \,{=}\, 1 - \sum \limits_{l = 0}^{i-1} \frac{e^{-\Delta(L_2^2 - L_1^2)\lambda} \left[\Delta(L_2^2 - L_1^2)\lambda \right]^l}{l!}$, the joint CDF in \eqref{eq:JCDF} is readily available. Taking derivative of the joint CDF, we obtain the joint PDF as follows
\begin{align} \label{eq:JPDF_derivation_step_1}
f_{d_k,d_i|\mathcal{S}_{K_2}} (r, r_i) &=\frac{\partial^2 F_{d_k,d_i|\mathcal{S}_{K_2}} (r, r_i)}{\partial r \partial r_i} \nonumber \\
&= - \frac{1}{\mathcal{C}}\frac{\partial}{\partial r_i}   \left\lbrace \frac{\partial}{\partial r} \left\lbrace \sum \limits_{l=k}^{i-1} \sum \limits_{l'=0}^{i-l-1} \frac{e^{-\Delta(r_i^2 - L_1^2)\lambda}\left[ \Delta(r^2 - L_1^2)\lambda \right]^l}{l!} \frac{\left[ \Delta(r_i^2 - r^2)\lambda \right]^{l'}}{l'!}\right\rbrace \right\rbrace
\nonumber \\
&= - \frac{1}{\mathcal{C}}\frac{\partial}{\partial r_i} \left\lbrace   \frac{(2\Delta \lambda r) \left[ \Delta(r^2 - L_1^2)\lambda \right]^{(k-1)}}{(k-1)!} e^{-\Delta(r_i^2 - L_1^2)\lambda} \sum \limits_{l'=0}^{i-k-1} \frac{\left[ \Delta(r_i^2 - r^2)\lambda \right]^{l'}}{l'!}\right\rbrace
\nonumber \\
&=\frac{(2\Delta \lambda r) }{\mathcal{C}}   \frac{\left[ \Delta(r^2 - L_1^2)\lambda \right]^{(k-1)}}{(k-1)!} (2\Delta \lambda r_i) e^{-\Delta(r_i^2 - L_1^2)\lambda} \frac{\left[ \Delta(r_i^2 - r^2)\lambda \right]^{(i-k-1)}}{(i-k-1)!}.  \;\; \IEEEQEDhere
\end{align}

\section{Asymptotic Analysis of Outage Probabilities} \label{app:Asymp_analysis_outage_j_i}
In this section, the approximated conditional outage probabilities from asymptotic analysis of \eqref{eq:cond_outage_j_sk1_3} and \eqref{eq:cond_outage_kn_sk2_3} are derived assuming 1) $2\Delta \rightarrow 0$, and 2) high SNR $\left( P_{\rm Tx}/N_0\rightarrow \infty \right)$.

\subsubsection{Asymptotic Behavior of Fej\'er Kernel}
As discussed in Section~\ref{sec:sumrate_noma} when $2\Delta\,{\rightarrow}\,0$ angular offset $|\overline{\theta}\,{-}\,\theta_k|\,{\rightarrow}\,0$. Hence, ${{\rm F}_M \left(\pi[\overline{\theta}\,{-}\,\theta_k]\right) }$ can be approximated as, \begin{align} \label{eq:fejer_kernel_1}
{{\rm F}_M \left(\pi[\overline{\theta}\,{-}\,\theta_k]\right) } \approx M \, {\rm sinc}^2 \left( \frac{\pi M(\overline{\theta} - \theta_k)}{2} \right).
\end{align}

Employing the facts 1) ${\rm sinc}(x) \approx 1-\frac{x^2}{6}$, and 2) $(1-x)^2 \approx 1-2x$ when $x\rightarrow 0$~\cite{Num_Analysis}, \eqref{eq:fejer_kernel_1} can be further approximated as
\begin{align} \label{eq:fejer_kernel_2}
{{\rm F}_M \left(\pi[\overline{\theta}\,{-}\,\theta_k]\right) } &\approx M \left( 1 - \frac{1}{6} \left(  \frac{\pi M(\overline{\theta} - \theta_k)}{2}  \right)^2 \right)^2 = M \left( 1 -  \frac{\pi^2 M^2(\overline{\theta} - \theta_k)^2}{24}   \right)^2 \nonumber \\
&\approx  M \left( 1 -  \frac{\pi^2 M^2(\overline{\theta} - \theta_k)^2}{12}   \right).
\end{align}

\subsubsection{Asymptotic Analysis of Conditional Outage Probability for $\mathcal{S}_{K_1}$} \label{sec:asy_ana_Sk1}
The exact analytical expression of the outage probability for this case ${\rm P}_{j|\mathcal{S}_{K_1}}^{o,\, 3}$ is given in \eqref{eq:cond_outage_j_sk1_3}.
Incorporating the asymptotic Fej\'er kernel from \eqref{eq:fejer_kernel_2} into \eqref{eq:cond_outage_j_sk1_3} assuming $2\Delta \rightarrow 0$, which yields
\begin{align} \label{eq:Outage_Sk1_2}
{\rm P}_{j|\mathcal{S}_{K_1}}^{o,\, 3} \approx \frac{1}{{\rm P} \{{\rm E}_3\}} \int\limits_{\overline{\theta}{-}\Delta}^{\overline{\theta}{+}\Delta} \int\limits_{l_1}^{l_2}  \left( 1 - \exp \left\lbrace \frac{-\eta_j{\textrm{PL} \left(\sqrt{r^2 + h^2}\right)}}{ {M \left( 1 -  \frac{\pi^2 M^2(\overline{\theta} - \theta_k)^2}{12}   \right)}} \right\rbrace  \right) \frac{f_{d_j|\mathcal{S}_{K_1}}(r) }{2\Delta} \dd r \dd \theta .
\end{align} Note that, $(1-x)^{-1}\approx 1+x$ when $x \rightarrow 0$. Employing this fact, outage probability in \eqref{eq:Outage_Sk1_2} is:
\small
\begin{align}
\label{eq:Outage_Sk1_3}
{\rm P}_{j|\mathcal{S}_{K_1}}^{o,\, 3} \approx \frac{1}{{\rm P} \{{\rm E}_3\}} \int\limits_{\overline{\theta}{-}\Delta}^{\overline{\theta}{+}\Delta} \int\limits_{l_1}^{l_2}  \left( 1 - \exp \left\lbrace -\frac{\eta_j}{M} {\textrm{PL} \left(\sqrt{r^2 + h^2}\right)} {\left( 1 +  \frac{\pi^2 M^2(\overline{\theta} - \theta_k)^2}{12}   \right)} \right\rbrace  \right) \frac{f_{d_j|\mathcal{S}_{K_1}}(r) }{2\Delta} \dd r \dd \theta .
\end{align} \normalsize
Assuming high SNR regime to further approximate \eqref{eq:Outage_Sk1_3}, we have $\eta_j \rightarrow 0$ when $P_{\rm Tx}/N_0 \rightarrow \infty$, since $\eta_j\,{=}\,\frac{\epsilon_j}{P_{\rm Tx}/N_0}$. Considering $e^x\approx 1+x$ when $x \rightarrow 0$, we can further elaborate \eqref{eq:Outage_Sk1_3} as follows \begin{align} \label{eq:Outage_Sk1_4}
{\rm P}_{j|\mathcal{S}_{K_1}}^{o,\, 3} &\approx  \frac{1}{{\rm P} \{{\rm E}_3\}} \int\limits_{\overline{\theta}{-}\Delta}^{\overline{\theta}{+}\Delta} \int\limits_{l_1}^{l_2}  \frac{\eta_j}{M} {\textrm{PL} \left(\sqrt{r^2 + h^2}\right)} {\left( 1 +  \frac{\pi^2 M^2(\overline{\theta} - \theta_k)^2}{12}   \right)}   \frac{f_{d_j|\mathcal{S}_{K_1}}(r) }{2\Delta} \dd r \dd \theta \nonumber \\
& \approx \frac{1}{{\rm P} \{{\rm E}_3\}}  \int\limits_{l_1}^{l_2}  \frac{\eta_j}{M} {\textrm{PL} \left(\sqrt{r^2 + h^2}\right)} f_{d_j|\mathcal{S}_{K_1}}(r) \dd r  \int\limits_{\overline{\theta}{-}\Delta}^{\overline{\theta}{+}\Delta} \frac{1}{2\Delta} {\left( 1 +  \frac{\pi^2 M^2(\overline{\theta} - \theta_k)^2}{12}   \right)}  \dd \theta ,
\end{align} where \begin{align}
\label{eq:asy_Outage_intergral_angle}
\int \limits_{\overline{\theta}{-}\Delta}^{\overline{\theta}{+}\Delta} \frac{1}{2\Delta} {\left( 1 +  \frac{\pi^2 M^2(\overline{\theta} - \theta_k)^2}{12}   \right)}  \dd \theta= \left(  1+\frac{\pi^2 M^2 \Delta^2}{36} \right).
\end{align} Denoting the integral over $r$ in \eqref{eq:Outage_Sk1_4} as $ \Psi_j = \int _{l_1}^{l_2}  \frac{{\textrm{PL} \left(\sqrt{r^2 + h^2}\right)} }{M} f_{d_j|\mathcal{S}_{K_1}}(r) \dd r$ and using the result in \eqref{eq:asy_Outage_intergral_angle}, the final approximated outage probability  can be given as:
\begin{align} \label{eq:Outage_Sk1_6}
{\rm P}_{j|\mathcal{S}_{K_1}}^{o,\, 3} \approx \frac{1}{{\rm P} \{{\rm E}_3\}}\left(  1+\frac{\pi^2 M^2 \Delta^2}{36} \right)\Psi_j \eta_j.
\end{align} 
\subsubsection{Asymptotic Analysis of Conditional Outage Probability for $\mathcal{S}_{K_2}$}
The exact analytical expression of the outage probability of the $k$-th user for this case ${\rm P}_{k|\mathcal{S}_{K_2}}^{{\rm o},\, n}$ is captured in \eqref{eq:cond_outage_kn_sk2_3}.
Incorporating $2\Delta \rightarrow 0$ and high SNR assumptions, \eqref{eq:cond_outage_kn_sk2_3} becomes \small
\begin{align} \label{eq:Outage_Sk2_2}
&{\rm P}_{k|\mathcal{S}_{K_2}}^{{\rm o},\, n} \approx  \frac{1}{{\rm P} \{{\rm E}_n\}} \int\limits_{\overline{\theta}{-}\Delta}^{\overline{\theta}{+}\Delta} \int\limits_{u_n}^{v_n} \int\limits_{a_n}^{b_n}  \frac{\eta_k^{(n)}}{M} {\textrm{PL}\left(\sqrt{\delta_{ki}\ell^2 \,{+}\, \delta_{kj}r^2 \,{+}\, h^2}\right)} {\left( 1 +  \frac{\pi^2 M^2(\overline{\theta} - \theta_k)^2}{12}   \right)}   \frac{f_{d_j,d_i|\mathcal{S}_{K_2}}(r,\ell)}{2\Delta} \dd r \dd \ell \dd \theta \nonumber \\
& \approx \frac{1}{{\rm P} \{{\rm E}_n\}}  \int\limits_{u_n}^{v_n} \int\limits_{a_n}^{b_n}  \frac{\eta_k^{(n)}}{M} {\textrm{PL}\left(\sqrt{\delta_{ki}\ell^2 \,{+}\, \delta_{kj}r^2 \,{+}\, h^2}\right)} f_{d_j,d_i|\mathcal{S}_{K_2}}(r,\ell) \dd r \dd \ell  \int\limits_{\overline{\theta}{-}\Delta}^{\overline{\theta}{+}\Delta} \frac{1}{2\Delta} {\left( 1 +  \frac{\pi^2 M^2(\overline{\theta} - \theta_k)^2}{12}   \right)}  \dd \theta .
\end{align}  \normalsize
Denoting the integral in \eqref{eq:Outage_Sk2_2} as $
\Psi_{ij} = \int _{u_n}^{v_n} \int_{a_n}^{b_n}  \frac{\eta_k^{(n)}}{M} {\textrm{PL}\left(\sqrt{\delta_{ki}\ell^2 \,{+}\, \delta_{kj}r^2 \,{+}\, h^2}\right)} f_{d_j,d_i|\mathcal{S}_{K_2}}(r,\ell) \dd r \dd \ell
$ and using the result in \eqref{eq:asy_Outage_intergral_angle}, the final approximated outage probability expression becomes
\begin{equation} \label{eq:Outage_Sk2_4}
{\rm P}_{k|\mathcal{S}_{K_2}}^{{\rm o},\, n} \approx \frac{1}{{\rm P} \{{\rm E}_n\}}\left(  1+\frac{\pi^2 M^2 \Delta^2}{36} \right)\Psi_{ij} \eta_k^{(n)}.
\IEEEQEDhereeqn\quad \end{equation} \vspace{-2em}
\bibliographystyle{IEEEtran}
\bibliography{Doc_Ref}
\end{document}